\documentclass[aps,prb,twocolumn,superscriptaddress,showpacs, final]{revtex4-2} 
\usepackage[normalem]{ulem}
\usepackage{amssymb}
\usepackage{suffix}
\usepackage{mathtools}
\usepackage[utf8]{inputenc}
\usepackage{booktabs}
\usepackage{cases}
\usepackage[multiple]{footmisc}
\usepackage{dcolumn}
\usepackage{color,soul}
\usepackage{rotating}
\usepackage{perpage}
\usepackage{xcolor}
\usepackage{soul}
\usepackage{tikz}
\usepackage[T1]{fontenc}
\usepackage{etoolbox}
\usepackage{graphics}
\usepackage{siunitx}
\usepackage{hyperref}
\usepackage{float}
\usepackage{multirow}
\usepackage{mathtools}
\usepackage{bm}
\usepackage{url}
\usepackage{tikz}
\usepackage{tikz-3dplot}
\usepackage [outdir=./]{epstopdf}
\usepackage{bbold}
\usepackage{changes}

\allowdisplaybreaks
\begin{document}
\title{Floquet-Engineered Chern Insulator in two-dimensional $d_{x^2-y^2}$-Wave Altermagnets}

\author{Hosein Cheraghchi}
\email{cheraghchi@du.ac.ir}
\address{School of Physics, Damghan University, P.O. Box 36716-41167, Damghan, Iran}

\date{\today}
\vspace{1cm}
\newbox\absbox
\begin{abstract}
We investigate Floquet-engineered topological phases in two-dimensional $d_{x^2-y^2}$-wave altermagnets irradiated by circularly polarized light in the off-resonant regime. These materials exhibit large momentum-dependent spin-splitting governed by distinctive magnetic symmetries. Using a lattice model combined with Floquet theory, we demonstrate that irradiation induces the light-tunable quantum anomalous Hall phases with the Chern numbers up to $\pm 3$. The resultant phase diagram is verified by calculating the anomalous Hall conductivity and also the edge modes inside the band gap of a nanoribbon version of the altermagnet. Our findings establish d-wave altermagnets as promising platforms for realizing nonequilibrium topological states of matter. 

The low-energy continuum limit of the lattice-based Floquet Hamiltonian results in a linear and higher-order-in-momentum spin–orbit couplings, and also a Zeeman-like magnetization, all arising from light-induced virtual photon processes. The resulting higher-order spin–orbit coupling generates the additional gapless Dirac points which, together with the high-symmetry gap-closings, yield enhanced Berry curvature and high Chern numbers. The light irradiation effectively breaks the static $d_{x^2 - y^2}$-wave magnetic symmetry mixing in an isotropic photo-induced $s$-wave correction.
\end{abstract}
\maketitle

\section{Introduction}
Altermagnets, recently identified as a third class of collinear magnetic materials, have attracted significant attention by combining key features of both ferromagnets and antiferromagnets~\cite{Smejkal,PRX22}. The alternating non-relativistic spin-splitting in altermagnets is comparable to the stray field in ferromagnets, leading to time-reversal symmetry breaking. However, similar to antiferromagnets, their zero net magnetic moment in real space provides an ideal platform for developing ultradense and ultrafast memory devices~\cite{Nat_review,alter,RMP1}. Altermagnetism has been experimentally observed in materials such as $RuO_2$ ~\cite{RuO1,RuO2,RuO3}({\it controversial}) and $MnTe$ ~\cite{MnTe1,MnTe2} and has been theoretically proposed for a table of candidate materials~\cite{alter}.

The quantum anomalous Hall effect (QAHE) which is a time-reversal-breaking response, provides dissipationless chiral edge states~\cite{haldane,RMP2}.
A non-trivial topological band structure with spin-splitting is required to achieve these chiral edge states. This condition is satisfied by magnetic adatoms doped into thin topological insulators which originally led to observation of QAHE~\cite{QAHE_chang,dabiri1,dabiri2}. 
However, QAHE at zero net magnetization is highly favorable especially in altermagnets which are promising for low-power quantum computing~\cite{tunableQAHE,RMP1,MnSi}. Chiral antiferromagnets~\cite{MnGe,MnSn,MnIr} in a kagome lattice with non-colinear or non-coplanar ordering exhibit a large anomalous Hall effect comparable to that of ferromagnetic metals. A framework of Berry-phase concepts arising from the chiral spin structure is considered as the origin of this effect.

Altermagnetism provides a promising platform for realizing QAHE. Altermagnetism lifts spin degeneracy through specific crystal symmetries, leading to spin-momentum locking and alternating spin-splitting in reciprocal space. These spin-splitting patterns are governed by the magnetic symmetries such as $d_{x^2-y^2}$, $g$, or $i$-wave types. Consequently, even with the collinear magnetic ordering, an altermagnet exhibits a k-space asymmetric band topology that generates non-zero Berry curvature. The QAHE has been theoretically proposed in several collinear magnetically compensated systems, including twisted bilayers of $Mn Bi_2 Te_4$~\cite{tunableQAHE}, monolayer CrO~\cite{QAHE1}, MoO~\cite{MoO} and $RbCr_4S_8$~\cite{RbCrS}.

\begin{figure}
\includegraphics[width= 0.8\linewidth]{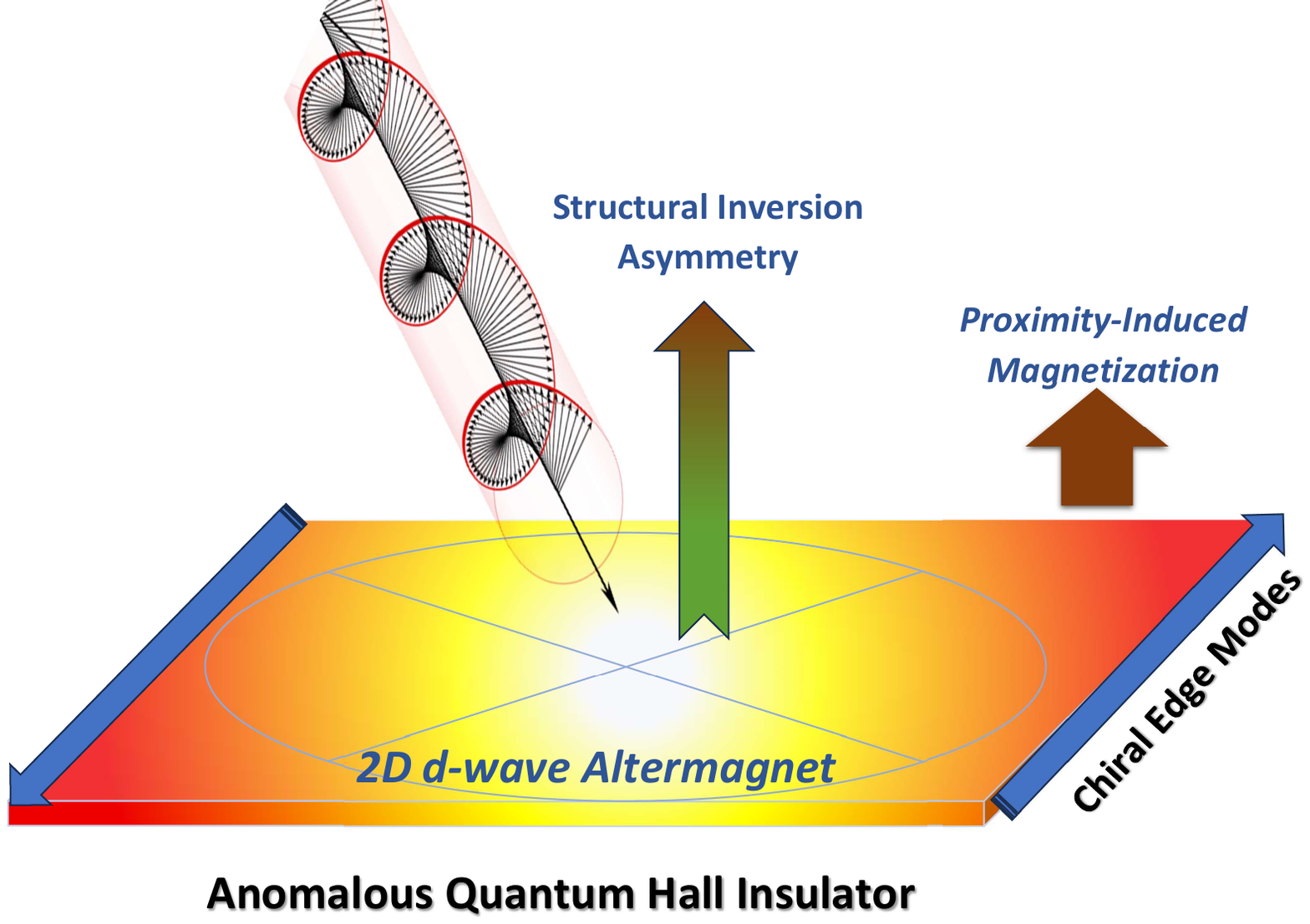}
\caption{Schematic illustration of a two-dimensional $d_{x^2-y^2}$-wave altermagnet irradiated by circularly polarized light. Structural inversion asymmetry, introduced by the substrate or interfacial proximity effects, generates a perpendicular electric field that induces Rashba spin-orbit coupling (RSOC). The interplay between RSOC, the $d_{x^2-y^2}$-wave altermagnetic exchange texture, and an out-of-plane magnetization leads to a quantum anomalous Hall insulating state with a high Chern number ($\mathcal{C}=\pm3$) under circularly polarized light irradiation. The out-of-plane magnetization is achievable by extrinsic exchange couplings induced by the proximity of a ferromagnet.}
\label{Cartoon}
\end{figure}

The interplay between altermagnetism and topology still remains largely unexplored. While spin-orbit coupling can generate topological gaps in specific symmetries as discussed in \cite{Smejkal}, the absence of a net magnetization imposes fundamental constraints on conventional topological mechanisms. Floquet engineering \cite{oka2009,rudner,gedik} provides a powerful pathway to overcome these limitations. Through periodic driving with circularly polarized light, time-reversal symmetry is broken dynamically~\cite{kitagawa,oka2009}. This induces artificial gauge fields capable of engineering Berry curvature and creating topological band structures inaccessible in equilibrium~\cite{rudner_review}.

In this work, we develop a comprehensive framework for Floquet topological states in light-irradiated $d_{x^2-y^2}$-wave altermagnets. Based on a systematic analysis of the lattice model, we show that circularly polarized light (CPL) in the off-resonant regime iduces non-trivial topological phases characterized by the Chern numbers $\mathcal{C}=\pm1,\pm2,\pm3$. To understand the origin of this high Chern numbers, we investigate the low-energy continuum limit that renormalizes the linear Rashba spin–orbit coupling (SOC) anisotropically and generates higher-order momentum-dependent spin-orbit terms. In this limit, a Zeeman-like out-of-plane magnetization is also induced by the CPL which originates from virtual two-photon processes. Moreover, the light irradiation adds an isotropic magnetic $s$-wave symmetry to the original static $d$-wave symmetry leading to Berry curvature redistribution. The anisotropic and higher-order light-induced SOC term leads to the gap openings at the new Dirac points as well as at the high-symmetry points giving rise to high Chern numbers in the phase diagram. Indeed, the Berry curvature distribution dynamically evolves with the light parameters giving rise to a sequence of QAH phases characterized by high Chern numbers. Our results demonstrate that $d_{x^2-y^2}$-wave altermagnets provide a platform for realizing light-induced quantum anomalous Hall effects.

The paper is organized as follows: Section~\ref{S2} introduces the lattice model and also investigates the topological properties of the lattice model with no irradiation. A brief review of the Floquet formalism is presented in Sec.~\ref{S3}. By applying Peierls substitution, the whole Floquet Hamiltonian is constructed by analytical calculation of Fourier components of time-periodic Hamiltonian in Sec.~\ref{S4}. The effective Floquet Hamiltonian is analytically derived in Sec.~\ref{S5} in the high frequency regime. The rich features of the phase diagram as a function of the light parameters is presented in Sec.~\ref{S6}. The resultant phase diagram is verified in Sec.~\ref{S7} by numerical calculation of the anomalous Hall conductivity and also the edge modes inside the band gap of a nanoribbon version of the altermagnet. Finally, we conclude in Sec.~\ref{S8}.

\section{Static Lattice Model}\label{S2}
A schematic illustration of a two-dimensional $d_{x^2-y^2}$-wave altermagnet irradiated by circularly polarized light is shown in Fig.~\ref{Cartoon}. The Rashba spin–orbit coupling (RSOC) can be induced either by interfacial proximity effects or by structural inversion asymmetry. In addition, an out-of-plane magnetization may be introduced on the altermagnetic plate. Such a Zeeman-like spin splitting can originate from several mechanisms. In altermagnets, which possess zero net intrinsic magnetization, an effective out-of-plane magnetization may be induced by an external magnetic field, by proximity-induced exchange coupling from an adjacent ferromagnet, or via the Edelstein effect in the presence of an in-plane electric field.

The effective two-band Hamiltonian for a $d_{x^2-y^2}$-wave altermagnet on a square lattice can be expressed in the compact form $H_{\text{static}}(\mathbf{k}) = \mathbf{d}^{\text{static}}(\mathbf{k}) \cdot \boldsymbol{\sigma}$, where $\boldsymbol{\sigma}=(\sigma_x,\sigma_y,\sigma_z)$ are the Pauli matrices and $\sigma_0$ denotes the identity matrix. The components of the vector $\mathbf{d}^{\text{static}}(\mathbf{k})$ are given by

\begin{equation}
\begin{aligned}
d_0^{\text{static}} &= 2t \bigl[2 - \cos(k_x a) - \cos(k_y a)\bigr], \\[2pt]
d_x^{\text{static}} &= -\lambda \sin(k_y a), d_y^{\text{static}}= \lambda \sin(k_x a), \\[2pt]
d_z^{\text{static}} &= 2J\bigl[\cos(k_y a) - \cos(k_x a)\bigr] + M_z .
\end{aligned}
\label{hamilk}
\end{equation}

Here, $a$ is the lattice constant. The term $d_0^{\text{static}}$ represents the spin‑independent kinetic energy on the square lattice. The intrinsic $d_{x^2-y^2}$‑wave altermagnetic exchange is encoded in $d_z^{\text{static}}$ and produces a momentum‑dependent spin splitting that alternates sign between the $k_x$ and $k_y$ directions. The Rashba spin–orbit coupling (RSOC) corresponds to the in‑plane components $d_x^{\text{static}}\sigma_x + d_y^{\text{static}}\sigma_y$. An additional uniform out‑of‑plane magnetization $M_z$ is included, which may arise from an external magnetic field, proximity coupling, or the Edelstein effect.

In an altermagnt, the inversion symmetry is broken, however, a combination of time-reversal and lattice symmetry (like a rotation or translation) is preserved which guarantees to have zero net magnetization. The explicit forms of inversion, time-reversal, and rotational symmetry operators, as well as their action on the Hamiltonian, are provided in Appendix~\ref{App:Symmetry}.

To determine the topological properties of the two-band Hamiltonian in Eq.~\ref{hamilk}, we examine the gap-closing conditions. The energy gap is given by $2\sqrt{\sum_{i=x,y,z} d_i^2}$. Thus, gap closing occurs when $d_i=0$. At the high-symmetry points $\Gamma$, $M(\pi,\pi)$, $X(\pi,0)$, and $Y(0,\pi)$ (in units of $a^{-1}$), the Rashba SOC terms vanish, so we have $d^{static}_x=d^{static}_y=0$. Consequently, gap closure occurs when the remaining component $d^{static}_z$ becomes zero at these points. The corresponding mass terms are: $M_z$ at $\Gamma$ and $M$, $M_z+4J$ at X, and $M_z-4J$ at Y points. The Chern number of the lower band, obtained by summing the contributions from these Dirac points, reads~\cite{Geometrical}
\begin{equation}
\mathcal{C}=\frac{1}{2}\bigl[ 2 \operatorname{sgn}(M_z) - \operatorname{sgn}(M_z+4J) - \operatorname{sgn}(M_z-4J) \bigr].
\label{static_chern}
\end{equation}

It follows that for zero external magnetization ($M_z=0$) the Chern number vanishes. For a finite $M_z$, the Chern number becomes $+1$ for $0 < M_z < 4J$ and $-1$ for $-4J < M_z < 0$, corresponding to a quantum anomalous Hall insulator (QAHI).

\begin{figure}
\includegraphics[width= 0.6\linewidth]{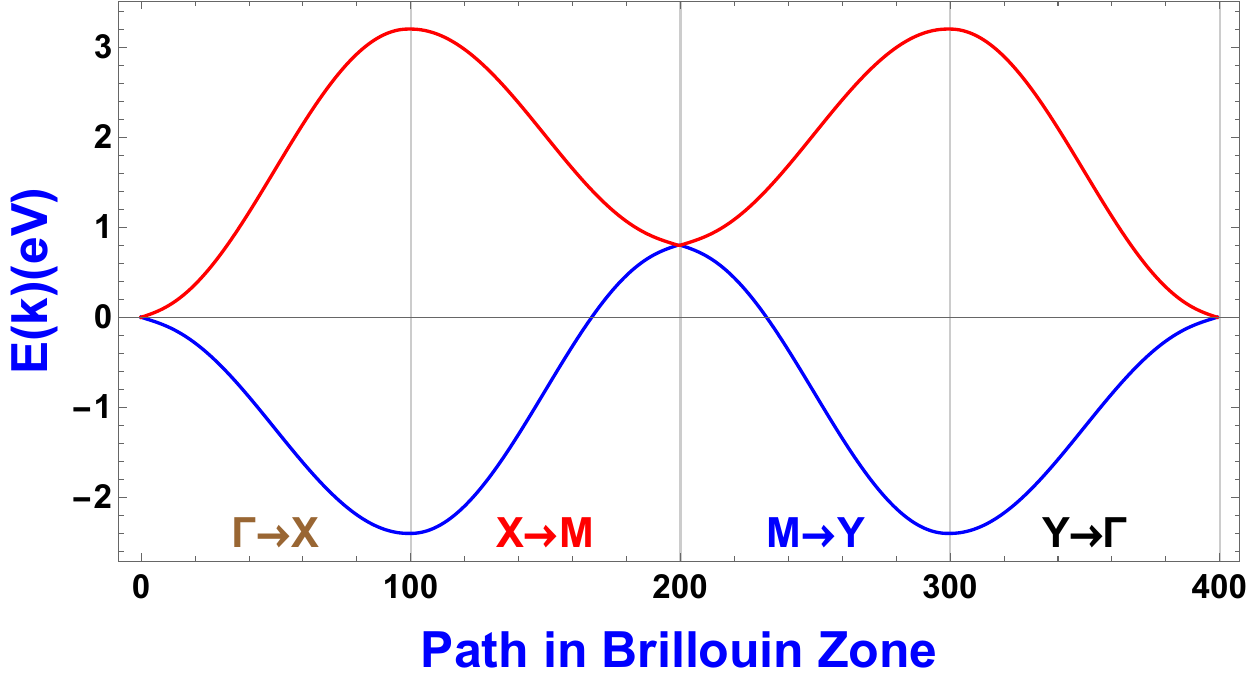}\includegraphics[width= 0.35\linewidth]{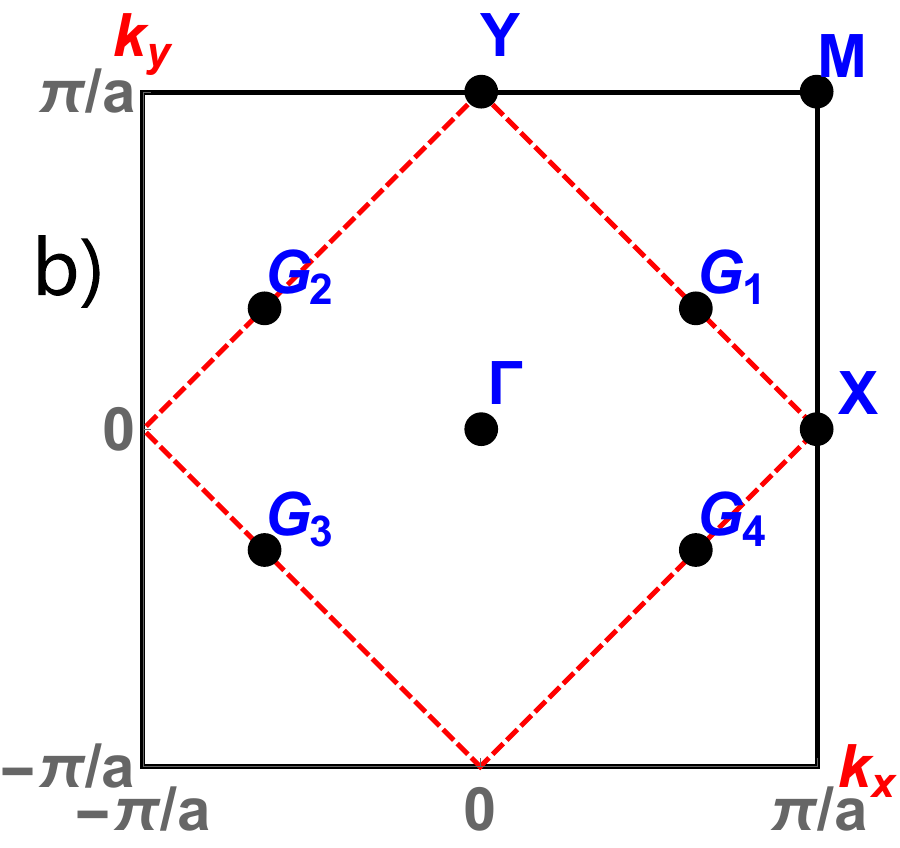}
\caption{The band structure of $d_{x^2-y^2}$- wave altermagnet in presence of Rashba spin-orbit coupling for the case of $M_z=0$ along a path including high symmetry points in the first Brillouin zone which is displayed in the right panel. Light illumination is turned off at this stage. Right panel: Four new gapless G-points are indicated in the first Brillouin zone appearing in $d_{x^2-y^2}$-wave altermagnets irradiated by cirucularly polarized light. We set Rashba spin-orbit coupling, $\lambda=0.3 ~\text{eV}$ and the altermagnetic spin splitting, $J=0.7 ~\text{eV}$.
 }
\label{Band_structure}
\end{figure}

The band structure of the static Hamiltonian in Eq.~\ref{hamilk} for $M_z=0$ is shown in Fig.~\ref{Band_structure}. The chosen path spans the high-symmetry points of the first Brillouin zone. In the absence of a uniform out-of-plane magnetization, the spectrum remains gapless at the $\Gamma$ and $M$ points. Since the altermagnetic exchange term is momentum dependent, it does not open a global band gap. A finite and uniform out-of-plane magnetization is therefore required to open a topological gap.

One can reconstruct the Hamiltonian represented in Eq.~\ref{hamilk} as the following 
\begin{equation}
\label{H_k}
H^{static}(\textbf{k})=T_0+T_x e^{ik_xa}+T_x^\dagger e^{-ik_x a}+T_y e^{ik_ya}+T_y^\dagger e^{-ik_y a},
\end{equation} 
where 
the onsite matrix $T_0$ and the hopping matrices along the $x,y$ directions $T_x$ , $T_y$ are defined as
\begin{equation}
\begin{aligned} 
T_0=&4 t \sigma_0+M_z \sigma_z\\
T_x=&-t \sigma_0-J\sigma_z+\frac{\lambda}{2i}\sigma_y\\
T_y=&-t \sigma_0+J\sigma_z-\frac{\lambda}{2i}\sigma_x.
\end{aligned}
\label{static_TB}
\end{equation}
Now, the real-space tight-binding form of Hamiltonian \ref{hamilk} reads
\begin{equation}
\label{TB_Hamiltonian}
H^{static}=\sum_{\bf r}^{}[ {\bf c}_{\bf r}^\dagger T_0{\bf c}_{\bf r}+(
{\bf c}^\dagger_{\bf r+a\hat{x}} T_x{\bf c}_{\bf r}+{\bf c}^\dagger_{\bf r+a\hat{y}} T_y{\bf c}_{\bf r}+h.c.)]
\end{equation} 
where ${\bf c}_{\bf r}^\dagger$ and ${\bf c}_{\bf r}$ are the creation and annihilation operators of electron at site $\bf{r}$. 

The low-energy effective Hamiltonian is derived by expanding Hamiltonian in Eq.~\ref{hamilk} around the $\Gamma$ point ($k_i a \ll 1$) by setting $\sin(k_i a) \approx k_i a$ and $\cos(k_i a) \approx 1 - (k_i a)^2/2$. The resulting low-energy Hamiltonian for the $d_{x^2-y^2}$-wave symmetry takes the form
\begin{equation}
\begin{aligned} 
\lim_{\bf{k} \to 0} H^{\text{static}}(\bf{k})&= t a^2[k_x^2+k_y^2]\sigma_0+J a^2 [k_x^2-k_y^2]\sigma_z\\
&+\lambda a [k_x \sigma_y-k_y \sigma_x]+M_z \sigma_z
\label{eq:hamilchiral} 
\end{aligned} 
\end{equation}

\section{Floquet Formalism}\label{S3}
To investigate time-periodic Hamiltonians, such as those describing irradiated systems, the Floquet formalism provides the appropriate framework for analyzing the system’s evolution at stroboscopic times~\cite{shirley}. In this approach, since energy is not conserved, instead of solving the time-periodic eigenvalue problem $H(t)=H(t+T)$ directly, one introduces a unitary time-evolution operator over one driving period,$U(T)=\mathcal{T}\exp \left[-i\int_{0}^{T} H(t),dt\right]$, where $\mathcal{T}$ denotes time ordering. It is always possible to define an effective Hermitian operator, known as the Floquet Hamiltonian $H_F$, such that $U(T)=\exp(-i H_F T)$.

The eigenvalues of $U(T)$, which are physically meaningful, take the form $\exp(-i\varepsilon_\alpha T)$, where $\varepsilon_\alpha$ denotes the quasienergy of the Floquet Hamiltonian $H_F$ associated with band index $\alpha$. Analogous to Bloch’s theorem for spatially periodic systems, a Bloch-like solution for time-periodic systems can be written as $e^{-i\varepsilon_\alpha t}|\psi_\alpha(t)\rangle$, where $|\psi_\alpha(t)\rangle$ is periodic in time with period $T$~\cite{gedik}. Consequently, the time-dependent Schroedinger equation can be recast as
\begin{equation}
\left[H(t)-i\frac{\partial}{\partial t}\right]|\psi_\alpha(t)\rangle
= \varepsilon_\alpha |\psi_\alpha(t)\rangle .
\end{equation}

This equation is explicitly time periodic and is therefore naturally suited for a Fourier expansion in time, leading to a description in quasienergy space. The Floquet Hamiltonian is then obtained from the Fourier-transformed Schroedinger equation,
\begin{equation}
\centering
\sum_{m'}[H^{(m-m')}-m\Omega\delta_{mm'}] |\psi^{m'}_\alpha\rangle=\varepsilon_\alpha |\psi^{m}_\alpha\rangle.
\label{inf}
\end{equation}
where the Fourier components of the time-periodic Hamiltonian are defined as
\begin{equation}
H^{(m)}=1/T \int_{0}^{T} H(t) e^{ i m |\Omega| t} dt .
\end{equation}
Here, $|\psi^{m}_{\alpha}\rangle$ denotes the $m^{th}$ Fourier component of the time-periodic wave function $|\psi_{\alpha}(t)\rangle$. The quasienergy spectrum is periodic, such that $\varepsilon_{\alpha} \equiv \varepsilon_{\alpha} + n\Omega$, where $\Omega=2\pi/T$ is the driving frequency. As a result, the quasienergy spectrum can be restricted to the first Floquet zone, $-\Omega/2 < \varepsilon_{\alpha} < \Omega/2$. Diagonalization of the Floquet Hamiltonian in Eq.~\eqref{inf} yields the quasienergy spectrum and the corresponding Floquet states, which enables the direct evaluation of topological invariants.

\section{Floquet Hamiltonian}\label{S4}
We now specialize the general Floquet formalism introduced in Sec.~\ref{S3} to the d-wave altermagnetic lattice model. The system is driven out of equilibrium by irradiation with circularly polarized light, which provides the vector potential periodic in time as ${\textbf A}(t)=A_0(\sin\Omega t,\cos\Omega t)$, where the sign of the light frequency $\Omega$ distinguishes right-handed (positive) and left-handed (negative) circular polarization. The coupling between light and matter is modeled via the Peierls substitution. Accordingly, the time-periodic Hamiltonian is obtained by applying $H({\textbf k})\rightarrow H({\textbf k}+\mathcal{A}(t))$ to the tight-binding Hamiltonian given in Eq.~\ref{H_k}, where $\mathcal{A}\equiv {e A/\hbar}$. This procedure is equivalent to a modification in the tight-binding model, $T_{\hat{r}} \rightarrow T_{\hat{r}} e^{\mathcal{A}(t).\hat{r}a}$, where $\hat{r}$ denotes to the nearest-neghibour directions on a square lattice along the $x$ and $y$ axes. Here, we assume a spacially uniform vector potential in real space. Within the Fourier representaion of the Schroedinger equation given in Eq.~\ref{inf}, the $n^{th}$ Fourier components of the time-periodic Hamiltonian $H^{n}$, is obtained as
\begin{equation}
\begin{aligned} 
&H^{(n)}=T_0 \mathcal{\delta}_{n,0} +\mathcal{J}_n(a\mathcal{A}_0)\large[ T_x e^{i k_x a}(-1)^n+T^\dagger_x e^{-i k_x a}\\
&+T_y e^{i k_y a}(i)^{n}+T^\dagger_y e^{-i k_y a}(-i)^{n}\large]\\
&=(4 t \sigma_0+M_z \sigma_z) \mathcal{\delta}_{n,0}+ \mathcal{J}_n(a\mathcal{A}_0)\large[ (i)^n \lambda \sin (k_x a+\frac{n\pi}{2}) \sigma_y\\
&-\lambda \sin(k_y a+\frac{n\pi}{2}) \sigma_x-2(t \sigma_0+J\sigma_z) (i)^n\cos(k_x a+\frac{n\pi}{2})\\
&+2(-t \sigma_0+J\sigma_z)\cos (k_y a+\frac{n\pi}{2}))\large]
\label{hf}
\end{aligned} 
\end{equation}
where $\mathcal{J}_n$ denotes the Bessel Function of the first kind. To derive $H^n$, we have used the integral identity $\frac{1}{T} \int_0^T e^{i a \mathcal{A}(t).\hat{r}}e^{in\Omega t}dt=\mathcal{J}_n(a\mathcal{A}_0) e^{in(\pi-\theta)}$, where the angle $\theta$ corresponds to the nearest-neighbour directions on a square lattice, $0,\pi/2,\pi$ and $3\pi/2$. The resultant Fourier component $H^n$ proves that the time-averaged hopping parameters of the square lattice are renormalized by the factor $\mathcal{J}_n(a\mathcal{A}_0) e^{in(\pi-\theta)}$relative to their static values. The full Floquet Hamiltonian is then constructed via the eigen-value equation in Eq.~\ref{inf}. 

\section{High frequency regime}\label{S5}
Although many Floquet sidebands emerge from the solution of Eq.~\ref{inf}, in the off-resonant regime the transition rate between the central Floquet band and higher-order sidebands is negligibly small. In this limit, an effective Floquet Hamiltonian can be obtained by projecting the full Floquet spectrum onto the first Floquet zone, $\varepsilon_{\alpha}\in [-\Omega/2,\Omega/2]$. The high-frequency regime, in simple terms, corresponds to a driving frequency much larger than the electronic bandwidth and a light intensity low enough to avoid resonant transitions. By using perturbation theory, it is straightfoward to show ~\cite{Bukov,eckart,bw} that the following series expansion in powers of the inverse frequency leads us to an effective Floquet Hamiltonian,
\begin{equation}
H^{\text{eff}}=H^{(0)}+\sum_{n=1}^{\infty}(n \Omega)^{-1}[H^{(-n)},H^{(+n)}]+O(1/(\Omega)^2)
\label{photon_H}
\end{equation}
As demonstrated in App.~\ref{FT}, the commutators of the even Fourier components vanish, i.e.,  $[H^{(-2m)},H^{(2m)}]=0$. In other words, the effective Hamiltonian can be written as:
\begin{equation}
H^{\text{eff}}=H^{(0)}+\frac{1}{\Omega}\sum_{m=0}^{\infty}\frac{[H^{(-(2m+1))},H^{(2m+1)}]}{2m+1}
\end{equation}
The resulting effective Hamiltonian is conveniently expressed in the form $H_{\text{eff}}=\textbf{d}^{\text{eff}}.\boldsymbol{\sigma}$, with

\begin{equation}
\begin{aligned}
d_0^{\text{eff}}&= 4 t -2t B\large[ \cos(k_x a)+\cos(k_y a)\large]\\
d_x^{\text{eff}}&=-\lambda\sin(k_y a)\large[ B+ C\cos(k_x a)\large]\\
d_y^{\text{eff}}&=  \lambda\sin(k_x a)\large[ B  -C \cos(k_y a)\large]\\
d_z^{\text{eff}}&=2J B\large[\cos(k_y a)-\cos(k_x a)\large]\\
&+J^{\prime} [\cos(k_x a) \cos(k_y a)]+M_z\\
\label{heff}
\end{aligned}
\end{equation}
Here, the dimensionless coefficients are defined as 
\[
 B=\mathcal{J}_0(a\mathcal{A}_0), C=\frac{8 J  }{\Omega}\mathcal{S}(a\mathcal{A}_0),
\]
while a {\it light-induced correction to the $d_{x^2-y^2}$-wave altermagnetic exchange energy} is given by 
\[
J^{\prime}=4\lambda^2 \mathcal{S}(a\mathcal{A}_0)/\Omega.
\]
The function $ \mathcal{S}$ is defined as 
\[
\mathcal{S}(x)=\sum_{m=0}^{\infty} (-1)^m (2m+1)^{(-1)} \mathcal{J}^2_{2m+1}(x)
\]
which is an oscillating function. In the static limit ($\mathcal{A}_0=0$), using $\mathcal{S}(0)=0$ and $\mathcal{J}_0(0)=1$ in Eq.~\ref{heff}, the static Hamiltonian in Eq.~\ref{hamilk} is straightforwardly recovered. The combined time-reversal and rotation symmetry that protect the static Hamiltonian is broken in the effective Floquet Hamiltonian giving rise to a light-induced out-of-plane magnetization. For detailed analysis of symmetries, please see App.~\ref{App:Symmetry}.

 To further substantiate the validity of the effective Floquet Hamiltonian, it is instructive to analyze its low-energy behavior in the weak-driving regime. In this limit, a continuum description becomes appropriate provided that the maximum energy modulation induced by the driving field between neighboring lattice sites ($a\mathcal{A}_0$) remains much smaller than the photon energy $\hbar\Omega$~\cite{Bukov}. This approach allows us to quantify how the spin-orbit coupling and the effective out-of-plane magnetization are renormalized by the periodic drive. We therefore examine the low-energy expansion of the effective Hamiltonian in Eq.~\ref{heff} around ${\mathbf k}=0$ in the weak-field regime.

\section{Continuum Limit}
\label{Continuum_Limit}
In the weak-field regime, the dominant contribution to the summation defining the $\mathcal{S}$-function arises from the lowest term $m=0$. Accordingly, one may approximate $\mathcal{S}(x) \simeq \mathcal{J}_1^2(x) \approx \frac{x^2}{4}$, while the zeroth-order Bessel function can be expanded as $\mathcal{J}_0(x) \simeq 1 - x^2/4$. Using these expansions, we derive the effective driven Hamiltonian in the high-frequency limit. The resulting expressions reproduce the continuum Hamiltonian reported previously in Refs.~\cite{Yarmohammadi,Ghorashi}, while additional correction terms reveal in our lattice model that are absent in their ${\bf k\cdot p}$ formulation. 
The effective Hamiltonian can be written as
\begin{equation}
\begin{aligned}
d_0^{\text{eff}} &= t \left( k_x^2 + k_y^2 + \mathcal{A}_0^2 \right) a^2 ,\\
d_x^{\text{eff}} &= -(\lambda + \lambda^{\prime}) k_y a
+ \frac{\lambda^{\prime}}{2} k_x^2 k_y a^3 ,\\
d_y^{\text{eff}} &= (\lambda - \lambda^{\prime}) k_x a
+ \frac{\lambda^{\prime}}{2} k_y^2 k_x a^3 ,\\
d_z^{\text{eff}} &= \left[ J (k_x^2 - k_y^2)
- \frac{J^{\prime}}{2} (k_x^2 + k_y^2) \right] a^2
+ M_{\text{light}} + M_z ,
\label{heff_low}
\end{aligned}
\end{equation}
where the light-induced spin-orbit couplings are linear and cubic in momentum and their SOC strength is scaled as
\[
\lambda^{\prime} = \frac{2 J J^{\prime}}{\lambda},
\]
where $J^{\prime} \simeq \frac{(\lambda a \mathcal{A}_0)^2}{\Omega}$, reflecting the contribution of virtual two-photon absorption and emission processes. The quantity
\[
M_{\text{light}} = J^{\prime}
\]
represents a light-induced Zeeman-like spin splitting.

We first focus on the effective out-of-plane spin splitting $d_z^{\text{eff}}$, which contains contributions associated with distinct magnetic symmetries as well as two momentum-independent magnetization terms. The term $J (k_x^2 - k_y^2)$ corresponds to the static altermagnetic exchange interaction, whose nodes lie along the diagonal line $k_x = k_y$. In contrast, the term proportional to $J^{\prime} (k_x^2 + k_y^2)$ provides an isotropic $k^2$ correction originating from virtual photon processes, and it remains finite even in the absence of the static exchange $J$.

As a consequence, CPL breaks the pure $d_{x^2-y^2}$-wave magnetic symmetry of the static altermagnet by admixing an effective $s$-wave component. This symmetry mixing lifts the symmetry-protected band degeneracies along the diagonal $k_x = k_y$ direction, opening additional gaps in the spectrum and generating new hot spots in the Berry curvature. Therefore, the fully dressed magnetic symmetry of irradiated altermagnets reflects the combined effects of the compensated magnetic order inherent to the static system and the symmetry breaking induced by periodic driving. In the high-frequency and weak-driving regime considered here, the induced correction remains perturbative, satisfying $J^{\prime} \ll J$.

In addition, a uniform and momentum-independent magnetization $M_{\text{light}}$ emerges in $d_z^{\text{eff}}$ as a direct consequence of CPL. More generally, in two-dimensional Rashba systems it has been shown that CPL generates an effective magnetic field through the Floquet commutator $[H_{-1}, H_{+1}]$. This induced magnetization is oriented along the propagation direction of the incident light~\cite{Sato}.

Finally, the off-diagonal components of the effective Hamiltonian in Eq.~\eqref{heff_low} reveal a renormalization of the Rashba SOC generated by the periodic drive~\cite{Yarmohammadi}. Indeed, CPL renormalizes the linear Rashba spin–orbit coupling, leading to anisotropic SOC strength along the $k_x$ and $k_y$ directions. This anisotropy, quantified by the parameter $\lambda^{\prime}$, is comparable in magnitude to the bare RSOC $\lambda$ and vanishes identically when $J = 0$, highlighting its intrinsic connection to altermagnetism.

In addition, a higher-order SOC term cubic in ${\bf k}$ is induced by the drive, distinct from conventional Rashba SOC and without counterpart in continuum ${\bf k.p}$ models~\cite{Yarmohammadi,Ghorashi}. The ${\bf k\cdot p}$ description, valid only around a single high-symmetry point (e.g. $\Gamma$), captures only the leading-order light-induced spin–orbit coupling. Thus, the continuum ${\bf k.p}$ models miss higher-order momentum-dependent terms and additional symmetry-related gap closing points. These higher-order SOC, together with the light-induced SOC anisotropy, substantially modify the Berry curvature distribution, creating new hotspots in the Brillouin zone. 
\begin{figure}
\includegraphics[width=0.55 \linewidth]{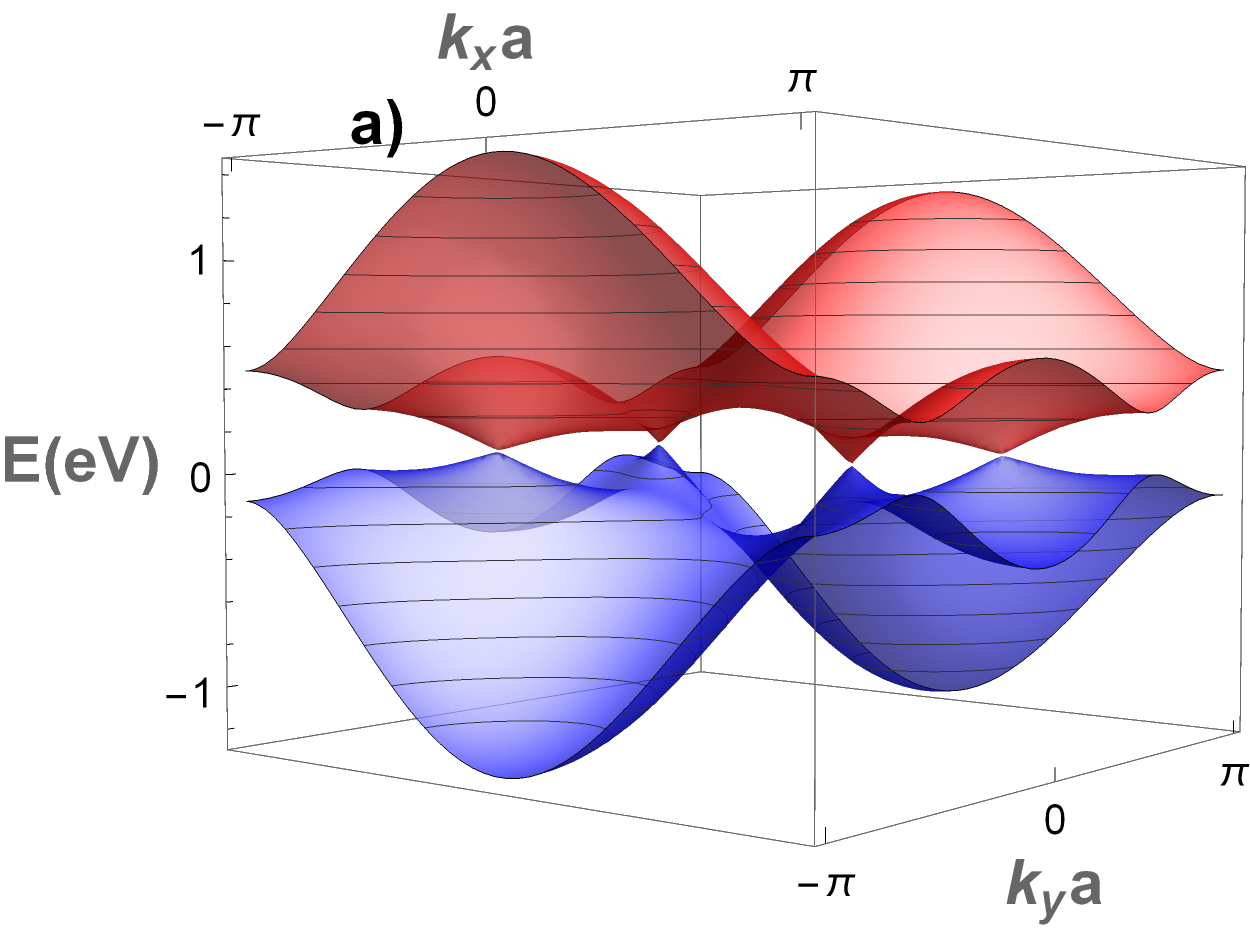}\includegraphics[width=0.4 \linewidth]{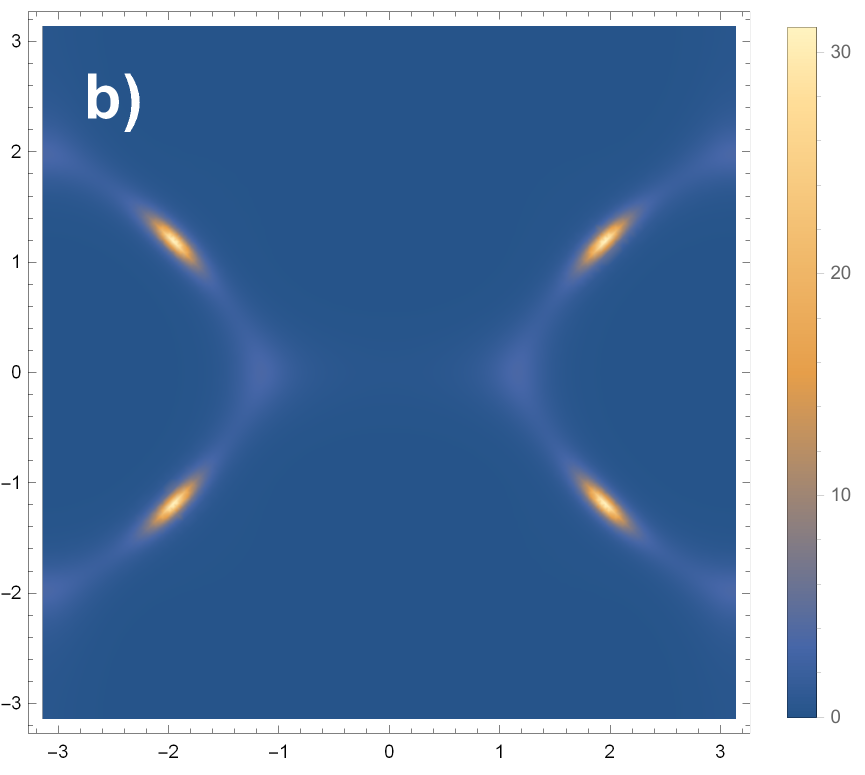}
\caption{a) The band spectrum of $d_{x^2-y^2}$- wave altermagnet irradiated by a light with the amplitude $A_0=0.4 (1/nm)$ and application of the external magnetization $M_z=-0.37 ~\text{eV}$. In this spectrum, the four gapless G-points are observed which are responsible for emerging of high Chern number in irradiated d-wave altermagnets. b) the feature of the Berry curvature close to the gapless condition of the G-points ($A_0=0.385 (1/nm)$) shown in the phase diagram, Fig.~\ref{phase_diagram_Mz}. An integral over the BZ results in high Chern number $\mathcal{C}=+3$ for this distribution of the Berry curvature. Here, system parameters are set as $\lambda=0.5 ~\text{eV}, J=1.0 ~\text{eV}$ and the light frequency $\Omega=5 ~\text{eV}$. }
\label{3DSpectrum}
\end{figure}

\begin{figure}
\includegraphics[width=0.8 \linewidth]{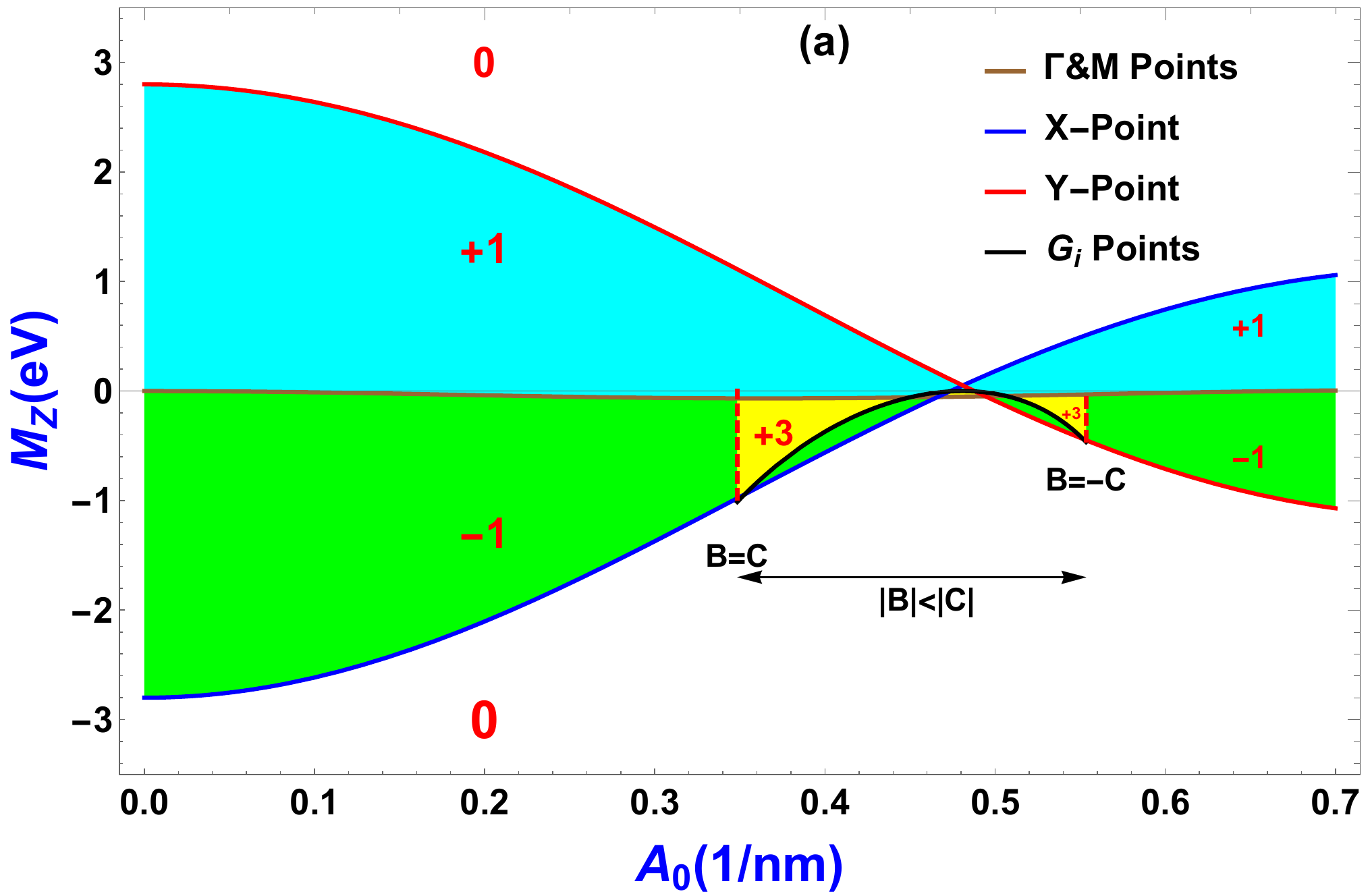} 
\includegraphics[width=0.8\linewidth]{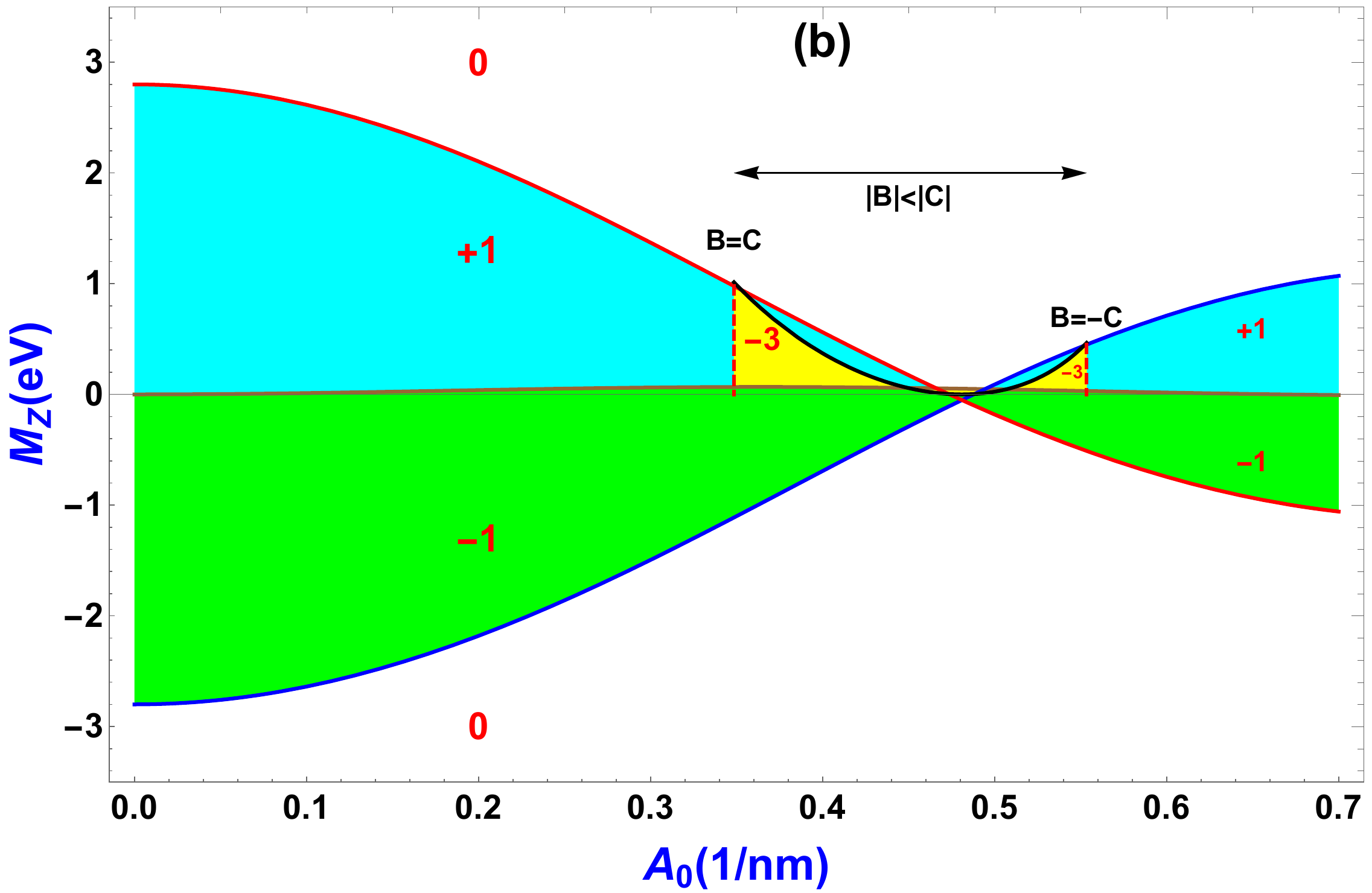}
\includegraphics[width=0.8\linewidth]{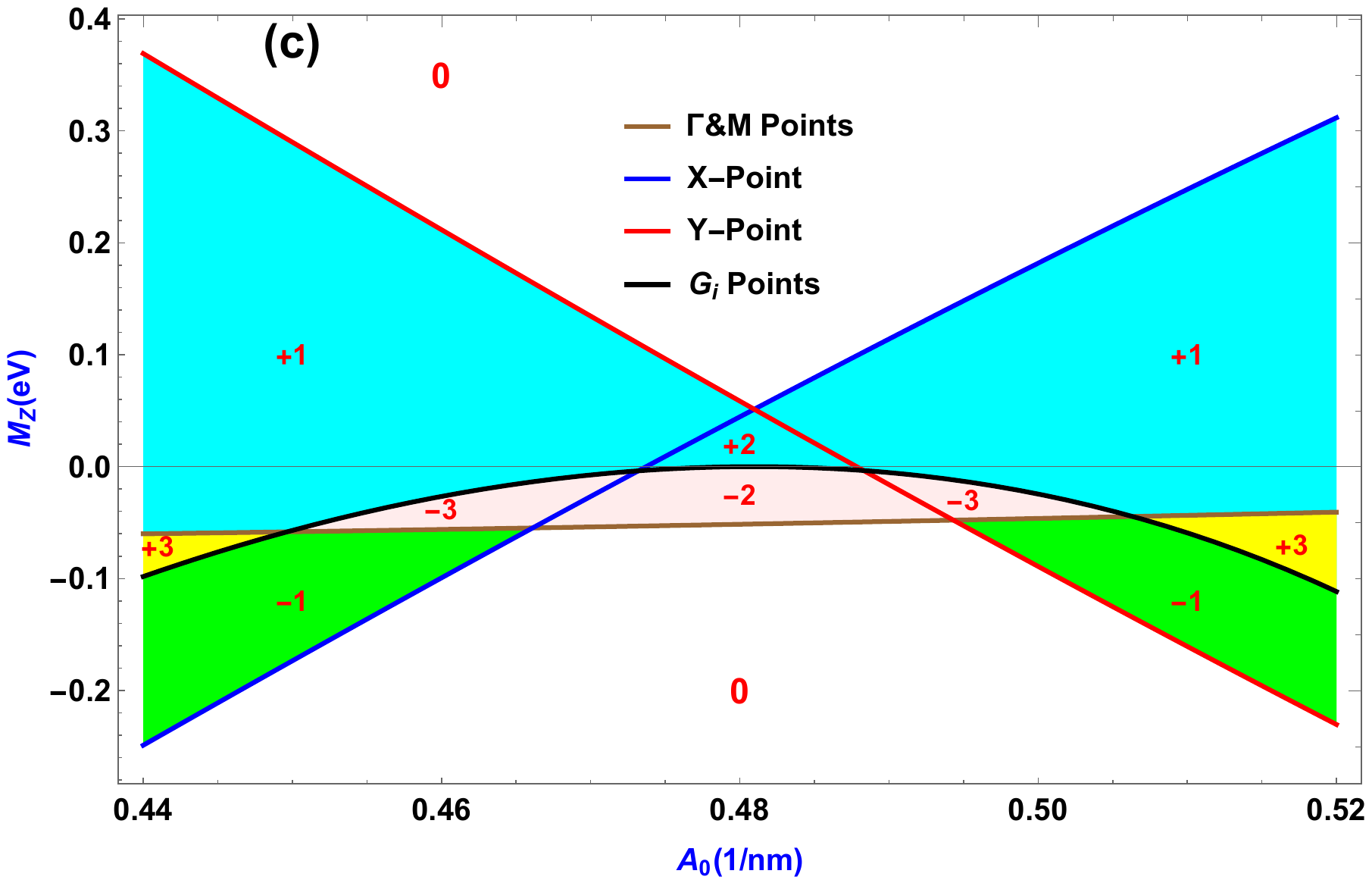}
\caption{(a) Phase diagram of a $d_{x^2-y^2}$-wave altermagnet under the right-handed circularly polarized light ($\mathrm{sgn}(\Omega)>0$) in the high-frequency regime, plotted as a function of the out-of-plane magnetization ($M_z$) and the light amplitude ($\mathcal{A}_0$). Red numbers denote the Chern number of the lower band of the effective off-resonant Hamiltonian [Eq.~\ref{heff}]. Gap-closing boundaries at the high-symmetry points ($\Gamma$), (M), (X), and (Y) separate phases with distinct topological invariants. At strong driving, additional gap closings at four momenta ($G_i$) ($i=1,\ldots,4$) occur in the regime $|B|<|C|$.
(b) Corresponding phase diagram for the left-handed circularly polarized light ($\mathrm{sgn}(\Omega)<0$). (c) Enlarged view of the $|B|<|C|$ region for the right-handed polarization, revealing multiple quantum anomalous Hall phases generated by the rearrangement of gap-closing boundaries. Parameters: $\lambda=0.5~\mathrm{eV}$, $J=0.7~\mathrm{eV}$, and $\Omega=5~\mathrm{eV}$.
 } 
\label{phase_diagram_Mz}
\end{figure}

\section{Phase Diagram}\label{S6}

In this section, we construct the topological phase diagram of the driven
$d_{x^2-y^2}$-wave altermagnet in the off-resonant regime.
The central idea is that topological phase transitions are separated by gap
closings of the effective Floquet Hamiltonian $H^{\text{eff}}$, and that the
Chern number of each insulating phase can be obtained by identifying the
location and chirality of the associated Dirac points.
For clarity, the detailed derivation of the gap-closing conditions is moved
to App.~\ref{gap_appendix}; here we summarize the results and focus on their
physical consequences.

\begin{figure}
\includegraphics[width=0.45\linewidth]{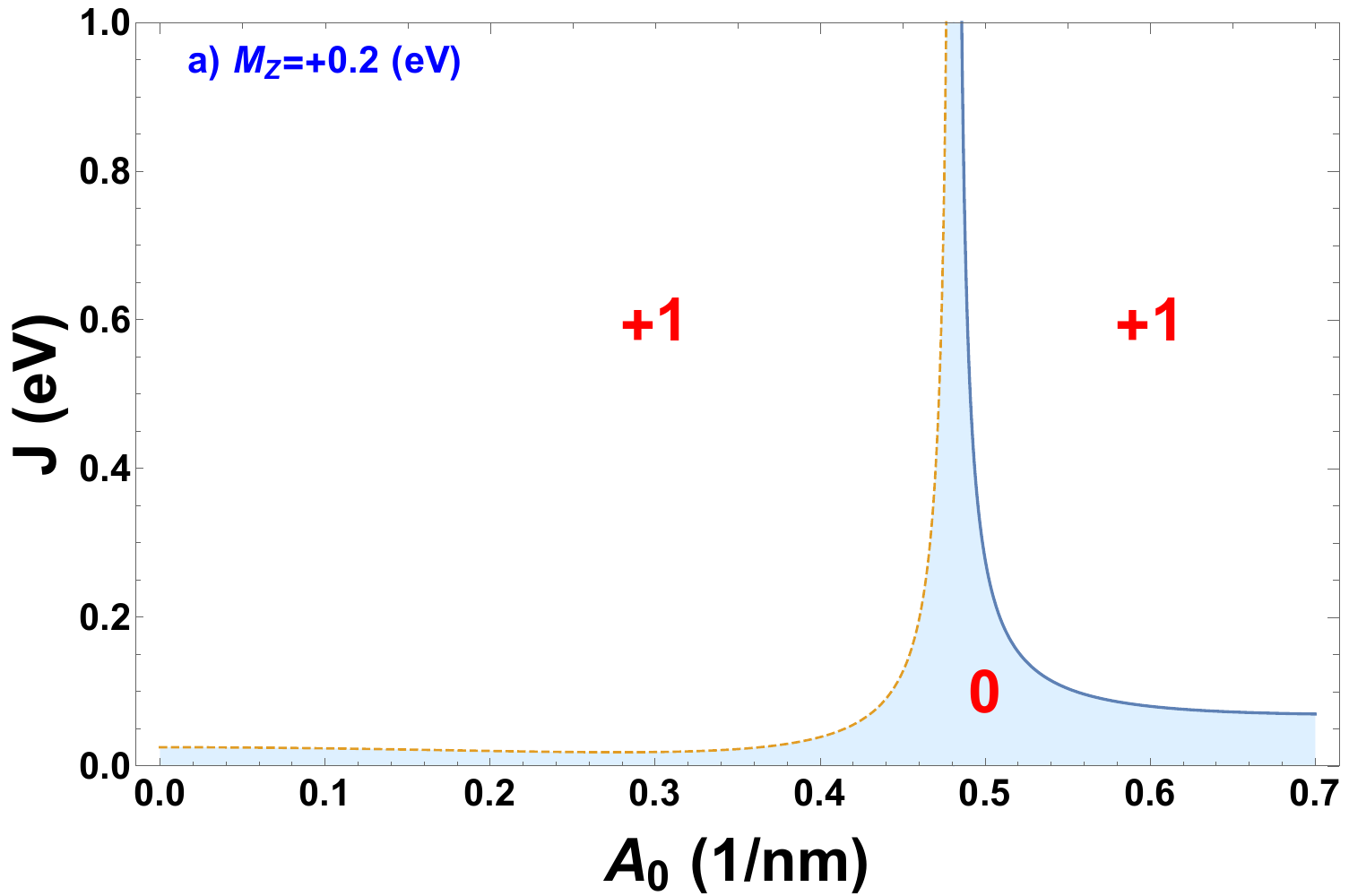} 
\includegraphics[width= 0.45\linewidth]{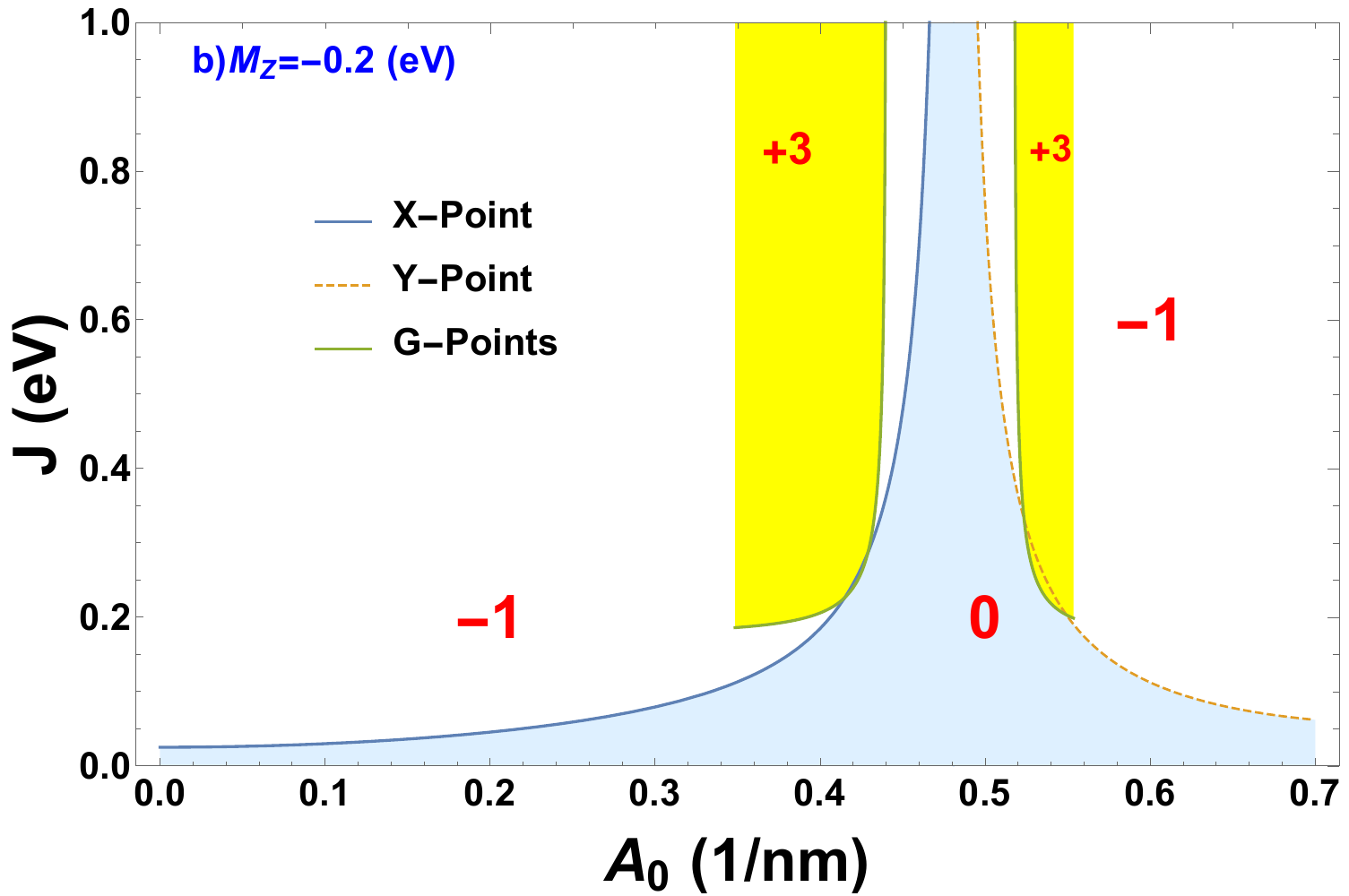} 
\caption{(a) Phase diagram of the $d_{x^2-y^2}$-wave altermagnet irradiated by the right-handed circularly polarized light, $sgn(\Omega)>0$, for the off-resonant regime shown as a function of the altermagnetc spin-splitting $J$ and the light amplitude $\mathcal{A}_0$. Panels correspond to (a) $M_z=+ 0.2 ~\text{eV}$ (b) $M_z=-0.2 ~\text{eV}$. Red numbers indicate the chern number for the lower band. Gap-closing boundaries at the $X, Y$ and $G$-points separates regions with distinct topological phases. Throughout the phase diagram, the mass term satisfies $sgn(d_z)>0 (<0)$ at the $\Gamma$ and $M$ points for $M_z= +0.2~\text{eV} (-0.2~\text{eV})$. Parameters: $\lambda=0.5$~\text{eV} and the light frequency $\Omega=5$~\text{eV}. } 
\label{phase_diagram_J}
\end{figure}

\subsection{Gap-closing points and their classification}

The gap is closed at the points where the condition $d^{\text{eff}}_x=d^{\text{eff}}_y=0$ is satisfied. Looking at Eq.~\ref{heff} leads simply us to the following conditions which must be simultaneously satisfied,
\begin{equation}
\left \{
\begin{aligned}
&\sin(k_y a)[B+C \cos(k_x a)]=0\\
&\sin(k_x a)[ B- C \cos(k_y a)]=0\\
\end{aligned}\right.
\label{gap_closing_conditions}
\end{equation}
Solving these conditions (see App.~\ref{gap_appendix}) reveals that in addition to the high-symmetry points, CPL generates four symmetry-related Dirac points located away from the
    high-symmetry lines, hereafter referred to as $G_i$ points. These points exist only in the parameter regime $|B|<|C|$ and are a purely
    Floquet-induced feature absent in the static altermagnet. As shown below, they play a crucial role in realizing high-Chern-number
    phases. The band spectrum in this gap-closing points is shown in Fig.~\ref{3DSpectrum}a.   

We note that the gap-closing conditions determining the phase boundaries are independent of the kinetic energy scale $t$, and therefore the resulting phase diagrams do not depend on $t$. Throughout our analysis, the bandwidth set by $t$ is chosen sufficiently small to ensure a global bulk gap for the corresponding parameter regimes. The lattice parameter $a$ of the square lattice is considered to be $5 nm$.

\subsection{Chirality and Chern number}

To determine the topological character of each gapped phase, we evaluate the
chirality of the Dirac points emerging at the gap-closing boundaries.
For a two-band model, the contribution of a Dirac point at momentum $\mathbf{k}$
to the Chern number is given by
$\mathcal{C}_{\mathbf{k}}=\frac{1}{2}\,\mathrm{sgn}(J_3 d_z)$,
where $J_3=(\partial_{k_x}\vec d\times\partial_{k_y}\vec d)_z$ is the
$z$-component of the Jacobian.
The explicit expressions for $J_3(\mathbf{k})$ are derived in
App.~\ref{Jacobian_app}; here we summarize the results:
\begin{equation}
\begin{aligned}
&J_3(\Gamma/M)=(\lambda a)^2(B^2-C^2),\\
&J_3(X/Y)=-(\lambda a)^2(B\mp C)^2<0,\\
&J_3(G_i)=(\lambda a)^2(C^2-B^2)^2>0 .
\end{aligned}
\label{Jacobian_summary}
\end{equation}
These expressions show that the chirality of the $X$ and $Y$ points is always
negative, whereas each $G_i$ point carries a positive topological charge.
The sign of the chirality at $\Gamma$ and $M$ depends on whether $|B|$ is larger
or smaller than $|C|$.

The total Chern number of the occupied band is obtained by summing the
contributions of all Dirac points encountered when tuning system parameters,
\begin{equation}
\mathcal{C}=\sum_{\mathbf{k}\in D_i}\mathcal{C}_{\mathbf{k}} .
\end{equation}

\begin{figure}
\includegraphics[width= 0.8\linewidth]{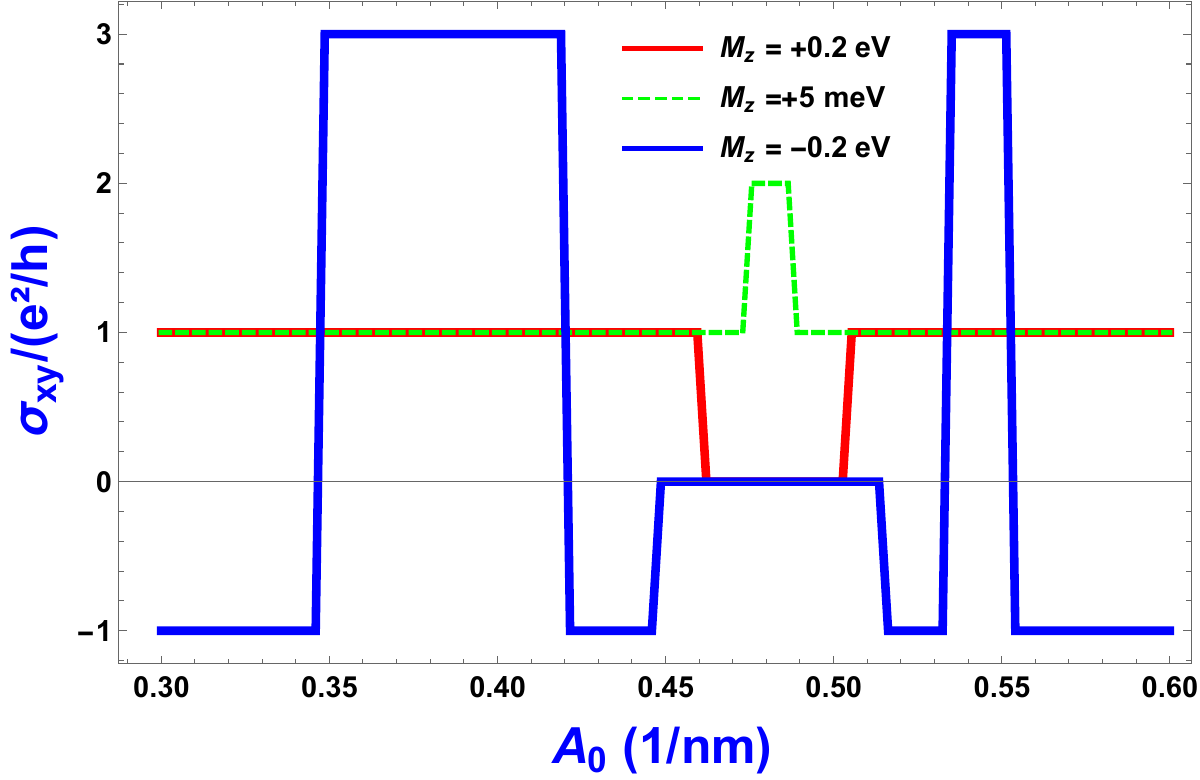} 
\caption{
Quantum anomalous Hall conductivity $\sigma_{xy}$ at zero temperature as a function of the light amplitude $\mathcal{A}_0$ for several values of the out-of-plane magnetization $M_z$, assumed to be uniform and $k$ independent. Results are shown for $M_z=+5~\mathrm{meV}$ and $M_z=\pm0.2~\mathrm{eV}$. The resulting Hall-plateau structure closely follows the topological phase boundaries shown in Figs.~\ref{phase_diagram_Mz} and \ref{phase_diagram_J}. Parameters are $J=0.7~\mathrm{eV}$, $\lambda=0.5~\mathrm{eV}$, and $\Omega=+5~\mathrm{eV}$.
}
\label{Hall_A0}
\end{figure}
\subsection{Phase diagrams and physical interpretation}

Using the above procedure, we now construct the phase diagrams in the
off-resonant regime. Figure~\ref{phase_diagram_Mz} shows the phase diagram as a function of the external magnetization $M_z$ and the light amplitude $\mathcal{A}_0$ for both right- and left-handed circularly polarized light. The Chern numbers of the occupied band, obtained analytically and independently confirmed using the Fukui–Hatsugai–Suzuki discretization scheme \cite{Discrete_BZ}.

For light intensities satisfying $|B|>|C|$, even a small external magnetization stabilizes a quantum anomalous Hall (QAH) phase with $|\mathcal{C}|=1$. In contrast, within the intensity window $|B|<|C|$, the emergence of the four $G$ points leads to high-Chern-number phases. For example, in Fig.~\ref{phase_diagram_Mz}(a), the total contribution
$$\mathcal{C}=(\mathcal{C}_{\Gamma}+\mathcal{C}_{M}+\mathcal{C}_X+\mathcal{C}_Y)+\sum_{i=1}^{4} \mathcal{C}_{G_i}=+3$$
originates from the combined effect of the high-symmetry and $G$ points. 

These large Chern numbers arise from light-induced higher-order spin–orbit coupling terms (discussed in Sec.~\ref{Continuum_Limit}), whose strength is controlled by the driving amplitude. Although the corresponding Dirac points exist only in a restricted parameter range, the window supporting high-Chern-number phases broadens significantly with increasing altermagnetic spin splitting $J$.

At low light intensities for right-handed circularly polarized light, the $\Gamma$ and $M$ points remain fully gapped when $M_z>0$. In contrast, for $M_z<0$ the gap closes at a critical magnetization $M_z^{cr.} \approx -2\lambda^2 x^2/\Omega$, leading to a topological phase transition at the $\Gamma$ and $M$ points, where $x=a \mathcal{A}_0$. Importantly, once the periodic drive is switched on, a quantum anomalous Hall (QAH) phase with $\mathcal{C}=\pm1$ emerges even in the absence of an external magnetization ($M_z=0$). This behavior is a purely Floquet-induced effect: according to Eq.~\ref{static_chern}, in the static (undriven) limit, the Chern number vanishes at $M_z=0$, and no QAH phase is realized. In particular, for zero external magnetization $M_z=0$, a Chern number $\mathcal{C}=1$ is postulated in Ref.~\cite{Ghorashi} for an off-resonant Floquet Hamiltonian derived from a ${\bf k\cdot p}$ model. However, since the ${\bf k\cdot p}$ model is valid only near a single high-symmetry point, the topology is controlled by a limited set of band inversions, yielding lower Chern numbers. In contrast, our lattice-based Floquet formulation represented in Eq.~\ref{heff} preserves the full crystalline and magnetic symmetries and explicitly incorporates higher-order Floquet-induced spin–orbit couplings, enabling high Chern numbers. 

Figure~\ref{phase_diagram_Mz}(b) shows the phase diagram for left-handed circularly polarized light. Reversing the light helicity does not qualitatively modify the overall structure of the phase diagram; however, it shifts the parameter window satisfying $|B|<|C|$. As a result, high-Chern-number phases are realized for opposite signs of the magnetization. In particular, a phase with $\mathcal{C}=-3$ appears for $M_z>0$, demonstrating that the sign of the Chern number can be controlled by the polarization of the driving field.

For completeness, Fig.~\ref{phase_diagram_J} shows the phase diagram as a function of the altermagnetic spin-splitting energy $J$ and the light amplitude $\mathcal{A}_0$ for two representative values of the out-of-plane magnetization, $M_z=\pm 0.2~\text{eV}$. Because the gap-closing condition at the $\Gamma$ and $M$ points, $M_z=- 4 \lambda^2 \mathcal{S}(a\mathcal{A}_0)/\Omega$, is independent of $J$, the corresponding gap-closing boundaries do not explicitly appear in this phase diagram. Nevertheless, their effect is implicitly reflected in the sign of the mass term: it is positive for $M_z=+0.2~\text{eV}$ and negative for $M_z=-0.2~\text{eV}$.

{\it Zoomed Phase Diagram}: In order to resolve fine topological structures that are not clearly visible in the global phase diagram, we provide in Fig.~\ref{phase_diagram_Mz}c a magnified view of the phase diagram within the parameter window $|B|<|C|$. This regime is of particular interest because it hosts light-induced gapless $G_i$ points and supports variety of the topological phases with high Chern numbers as well as an intermediate Chern number $\mathcal{C}=\pm 2$. The ordering of gap-closing boundaries directly modifies the accumulation of Dirac-point chiralities and, consequently, the evolution of the Chern number across the phase diagram.  For example, at a representative light amplitude $\mathcal{A}_0=0.35~\mathrm{nm}^{-1}$ [see Fig.~\ref{phase_diagram_Mz}(a)], the sequence of gap closings with increasing magnetization is $X \rightarrow G_i \rightarrow (\Gamma,M) \rightarrow Y$. In contrast, at a larger amplitude $\mathcal{A}_0=0.475~\mathrm{nm}^{-1}$, the zoomed plot in Fig.~\ref{phase_diagram_Mz}c shows that the ordering changes to $(\Gamma,M) \rightarrow G_i \rightarrow X \rightarrow Y$.

The evolution of the Chern number can be further understood by examining the Berry curvature of the Floquet bands. Light irradiation dynamically reshapes the Berry curvature distribution, generating pronounced hot spots around the gapless $G_i$ points. Each of these points contributes $\pm 1/2$ to the total Chern number, yielding an overall contribution of $\pm 2$ from the four $G_i$ points. Figure~\ref{3DSpectrum}(b) illustrates the Berry curvature profile near the gapless $G$ points, explicitly demonstrating their dominant role in stabilizing high-Chern-number topological phases.

\begin{figure}
\includegraphics[width= 0.8\linewidth]{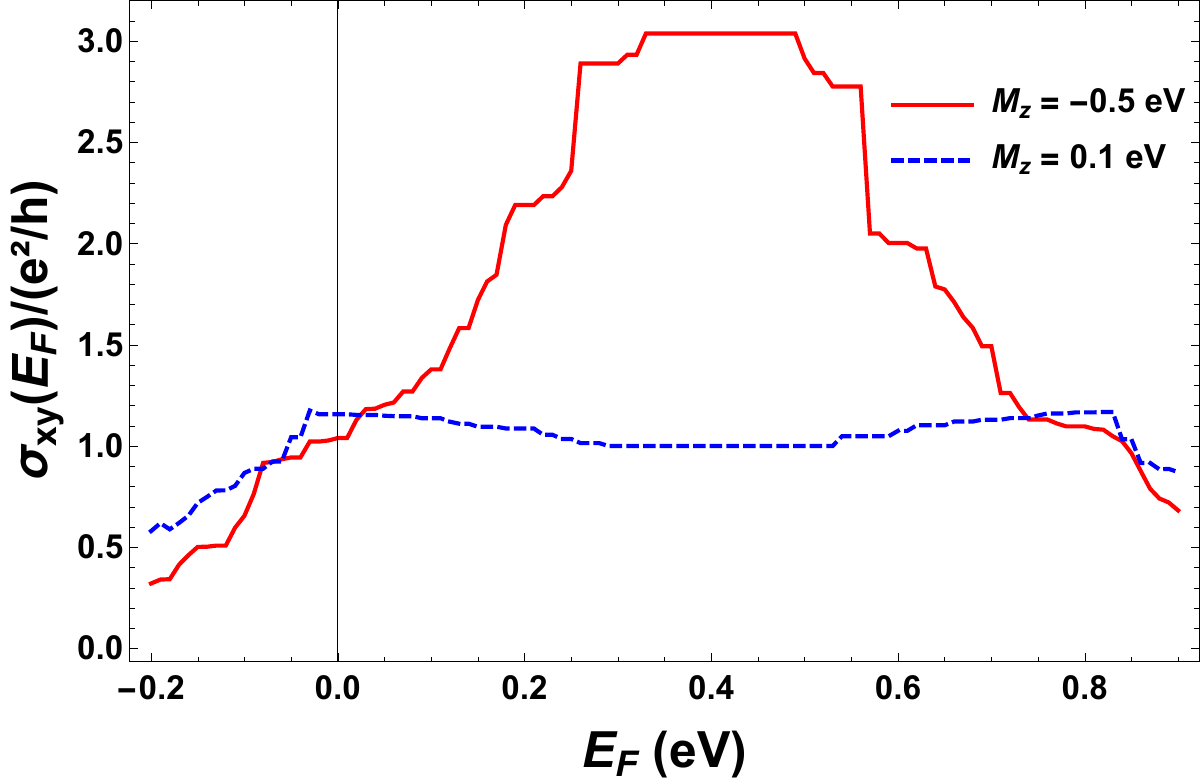} 
\caption{Quantum anomalous Hall conductivity $\sigma_{xy}$ as a function of the Fermi energy, $E_f$ for $M_z= -0.5, +0.1$~\text{eV}. Here, the light-altermagnet coupling is chosen to be $\mathcal{A}_0^2/\Omega=0.045$~\text{$nm^{-2}.eV^{-1}$}. Parameters: $J=1.0$~\text{eV}, $\lambda=0.6$~\text{eV}. } 
\label{Hall_energy}
\end{figure}
\begin{figure}
\includegraphics[width=0.7 \linewidth]{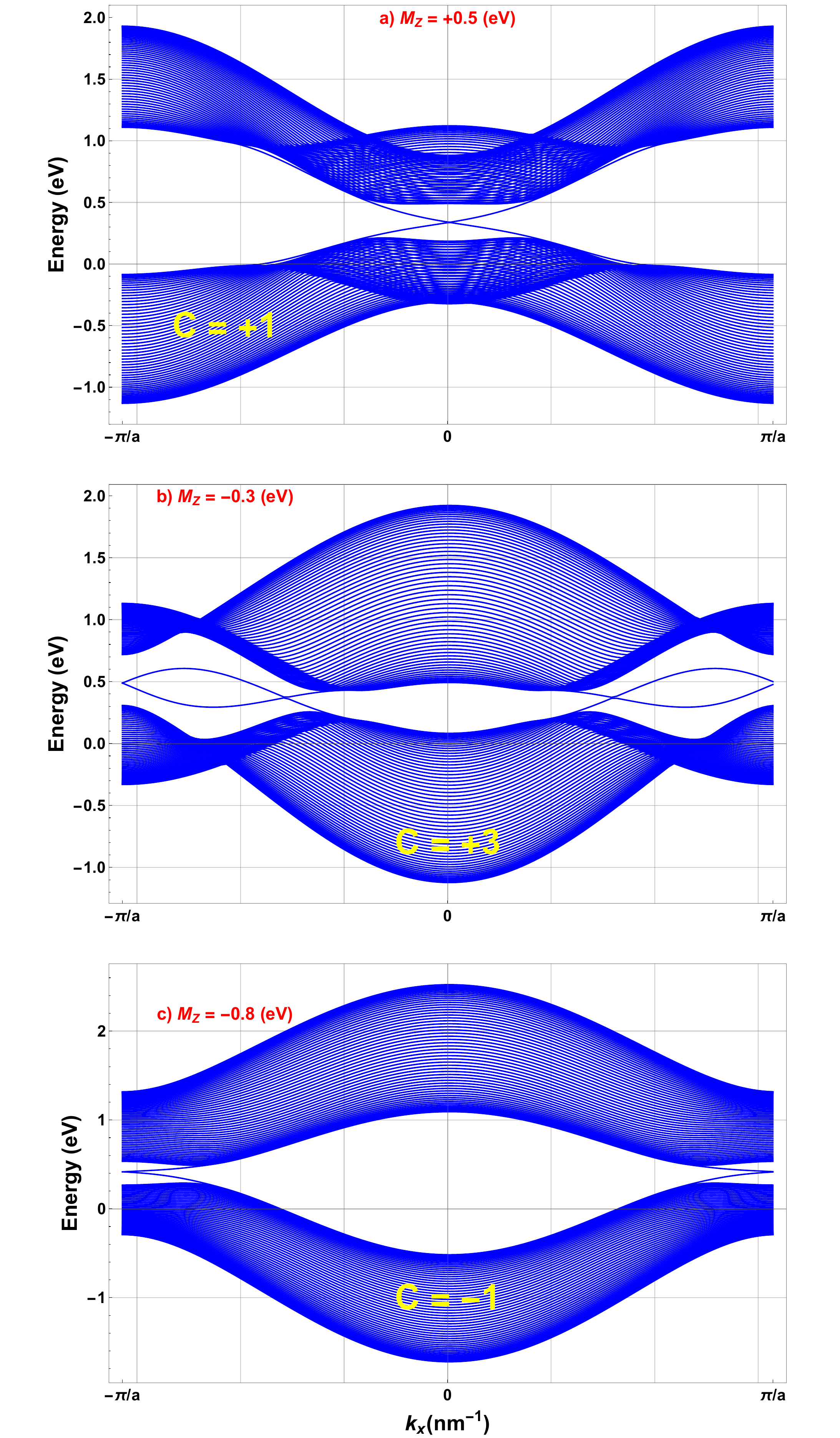} 
\caption{
Band structure of a nanoribbon of a $d_{x^2-y^2}$-wave altermagnet driven by right-handed circularly polarized light for out-of-plane magnetizations $M_z=0.5$, $-0.3$, and $-0.8~\mathrm{eV}$. The nanoribbon is oriented along the $x$ axis, and the energy spectrum is plotted as a function of $k_x$. Yellow numbers indicate the Chern number of the occupied bands. Parameters are $J=1.0~\mathrm{eV}$, $\lambda=0.6~\mathrm{eV}$, $\Omega=5.0~\mathrm{eV}$, and the light amplitude $\mathcal{A}_0=0.38~\mathrm{nm^{-1}}$.
}
\label{nanoribbon}
\end{figure}

In summary, the phase diagrams demonstrate that Floquet engineering provides a
powerful route to realizing QAH phases with large and tunable Chern numbers in
altermagnetic systems.
Importantly, these phases persist over an extended range of light amplitudes
and exchange energies, indicating that high-Chern-number QAH states should be
experimentally accessible.

\section{Verification of the Phase Diagram}\label{S7}
To confirm the accuracy and relevance of the resultant phase diagram, we try to verify this by using complementary approaches. The Chern numbers were computed by Fukui–Hatsugai–Suzuki lattice discretization method\cite{Discrete_BZ}. For the second verification, we evaluate Hall conductivty as a function of the light amplitude and energy by complying the Kubo formula for an insulator and at zero temperature.
\begin{equation}
\sigma_{xy} = 
\frac{e^2}{\hbar}
\sum_{n \in \text{occ}}
\int_{\mathrm{BZ}}
\frac{d^2\mathbf{k}}{(2\pi)^2} \,
\Omega_{n}(\mathbf{k}),
\label{eq:sigma_xy_berry}
\end{equation}
where $\Omega_{n}(\mathbf{k})$ is the Berry curvature of the $n^{th}$ band.
\begin{equation}
\Omega_{n}(\mathbf{k}) = 
-2\,\mathrm{Im}
\sum_{m \neq n}
\frac{
\langle n, \mathbf{k} | \hat{v}_x | m, \mathbf{k} \rangle
\langle m, \mathbf{k} | \hat{v}_y | n, \mathbf{k} \rangle
}{
\left(E_{m,\mathbf{k}} - E_{n,\mathbf{k}}\right)^2}.
\label{eq:berry_curvature}
\end{equation}
where $E_{n,\mathbf{k}}$ is the $n^{th}$ eigenvalue of the effective Floquet Hamiltonian, Eq.~\ref{heff}. Here, $v_{x}=\frac{1}{\hbar}\frac{\partial H_{\text{eff}}}{\partial k_x}$ is the velocity operator along the x-direction.

Figure~\ref{Hall_A0} shows anomalous Hall conductivity as a function of the light amplitude for different values of the magnetization, $M_z = 5~\text{meV}, \pm 0.2~\text{eV}$. The resultant Hall plateau structure closely follows the topological phases presented in Figs.~\ref{phase_diagram_Mz} and~\ref{phase_diagram_J}. Anomalous Hall conductivity which takes quantized values can be measured experimentally by using ultra-short high-frequency laser pulses. Interestingly, if the magnetization is small, as shown in Fig.~\ref{phase_diagram_Mz} c, for $0<M_z <20\text{meV}$, by turning the light on, one can observe a QAH phase whose Chern number evolves as $1 \rightarrow 2$. Such a small residual magnetization has been experimentally observed in the spontaneous anomalous Hall response of the $d$-wave altermagnet $Mn_5Si_3$, which originates from incomplete real-space magnetic compensation~\cite{MnSi}.

The QAH phase with high Chern number $+3$ manifests itself in the Hall conductivity as a function of the Fermi energy at zero temperature as well. Fig.~\ref{Hall_energy} indicates that in the band gap of the irradiated $d_{x^2-y^2}$-wave altermagnet, the anomalous Hall conductivity is quantized demonstrating the Chern number $\mathcal{C}=+3$ for $M_z=-0.5$~\text{eV}. A plateau is also observed in $\sigma_{xy}=+e^2/h$ for $M_z=0.1$~\text{eV} as long as the Fermi energy lies in the energy gap.

For the sake of compeleteness, we present in Fig.~\ref{nanoribbon}, the band structure of a nanoribbon version of the irradiated $d_{x^2-y^2}$-wave altermagnet laying along the x-direction for three representative values of the magnetization $M_z$ at the light-altermagnet coupling $\mathcal{A}_0^2/\Omega$=0.029~\text{ $nm^{-2}. eV^{-1}$}. For plotting the band structure displayed in Fig.~\ref{nanoribbon}a, the parameters are chosen such that a Chern insulator is yielded with $\mathcal{C}=+1$. As seen in Fig.~\ref{nanoribbon}a, the helical edge modes appeared within the band gap occur in the $\Gamma$ point. Indeed, confirming by the Berry curvature, these edge modes appear around the Y-points projecting along the nanoribbon direction, the x-axis. 

The intriguing band structure is dispalyed in Fig.~\ref{nanoribbon}b which is a clear evidence of the edge modes formed around the gapless G-points as the origin of high Chern number, $\mathcal{C}=+3$. In this case, there are three Dirac points; one occurs in the Brillouin boundary and two others are located symmetrically at $k_x=\pm \arccos(-B/C)/a$ which are the projection of the G-points on the x-axis. The last band structure is chosen such that $\mathcal{C}=-1$, and the edge modes appear at the Brillouin boundary. By looking at the Berry curvature, it is revealed that these edge modes are formed around the X-points projecting along the x-axis. To plot these band structures, we have used a tight-binding version of the effective Floquet Hamiltonian represented in Eq.~\ref{heff} whose tight-binding parameters are given in App.~\ref{TB}.

Finally, in App.~\ref{Pi_gap}, it is shown that the bulk-edge correspondence at the driving frequency used for plotting Fig.~\ref{nanoribbon} is not 
anomalous and reduces to the conventional form. It is well-known that aomalous edge states appear even when the Chern numbers of
all Floquet bands vanish \cite{rudner}. This phenomenon is associated with
edge modes traversing the quasienergy $\pi$ gap and cannot be captured by an
effective static Hamiltonian. However, it is numerically shown  in Fig.~\ref{Anomalous_edge} that no chiral edge states appear inside the $\pi$ quasienergy gap, and
all edge modes are confined to the zero quasienergy gap.

\section{Conclusion}\label{S8}
We have presented a systematic study on the topological properties of a lattice model describing a two-dimensional $d_{x^2-y^2}$-wave altermagnet irradiated by circularly polarized light in the high frequency, off-resonant regime. Considering an external, uniform and out-of-plane magnetization $M_z$, a Floquet phase diagram is represented in this work which exhibits a rich structure of the topological phases as function of the magnetization $M_z$ and the light amplitudes $\mathcal{A}_0$, including quantum anomalous Hall phases with the Chern numbers $\pm 1, \pm 2,\pm 3$. The origin of the high Chern numbers lie in the emergence of the additional light-induced Dirac points $G_i$ whose Berry curvature contribution is responsible for high Chern number added to those of the high-symmetry points.

To understand the origin of these new light-induced Dirac points, we investigate the lattic-based Floquet Hamiltonian in the low-energy continuum limit. In this limit, it is shown that the light vector potential anisotropically renormalizes the Rashba spin-orbit coupling (SOC) and also induces higher order-in-momentum SOC characterizing by the irradiated d-wave altermagnetism. Moreover, a Zeeman-like out-of-plane magnetization is induced by the light that scales with the Rashba SOC. The SOC corrections in competition with the originally linear term of Rashba SOC, produces additional Dirac points. The enhanced winding of the Floquet wavefunctions, arising from the dynamically generated anisotropic linear- and higher order-in-momentum SOC, underlies the observed topological complexity. Furthermore, in the low-energy limit, the light irradiation effectively breaks the original  $d_{x^2 - y^2}$ magnetic symmetry mixing in an isotropic $s$-wave correction proportional to $J^{\prime}k^2$ originated solely from the light-induced virtual photon processes. 


\appendix
\section{Symmetry analysis}
\label{App:Symmetry}

In this Appendix, we summarize the symmetry properties of the static Hamiltonian [Eq.~\ref{hamilk}] and the effective off-resonant Floquet Hamiltonian [Eq.~\ref{heff}] in the absence of external magnetization ($M_z=0$). We focus on time-reversal, inversion, rotational, and combined symmetries that are relevant for the emergence of topological phases~\cite{PRX22,alter}.

\subsection{Static Hamiltonian}

For spin-$1/2$ electrons, the time-reversal operator is $\Theta=i\sigma_y\mathcal{K}$, with $\mathcal{K}$ denoting complex conjugation. Under $\Theta$, momentum and spin transform as $\mathbf{k}\to-\mathbf{k}$ and $\boldsymbol{\sigma}\to-\boldsymbol{\sigma}$. Acting on the static Hamiltonian, the altermagnetic exchange term transforms as
\begin{equation}
\Theta\,\Delta_{\mathrm{ALT}}(\mathbf{k})\sigma_z\,\Theta^{-1}
= -\Delta_{\mathrm{ALT}}(-\mathbf{k})\sigma_z \neq \Delta_{\mathrm{ALT}}(\mathbf{k})\sigma_z,
\end{equation}
demonstrating explicit breaking of time-reversal symmetry by the momentum-dependent altermagnetic spin splitting. By contrast, the Rashba spin--orbit coupling (RSOC) term remains time-reversal invariant.

Inversion symmetry is generated by $\mathcal{P}:\mathbf{k}\to-\mathbf{k}$, leaving spin unchanged. While the kinetic energy and altermagnetic exchange terms are even under inversion, the RSOC term is odd, leading to
\begin{equation}
\mathcal{P}H^{\mathrm{static}}(\mathbf{k})\mathcal{P}^{-1}\neq H^{\mathrm{static}}(-\mathbf{k}),
\end{equation}
and hence inversion symmetry is broken. This broken inversion symmetry allows for a finite Berry curvature and anomalous Hall responses.

We next consider four-fold rotational symmetry about the $z$ axis. The four-fold rotation operator $C_{4z}$ contains two parts; the spatial rotaion acts as $(k_x,k_y)\to(-k_y,k_x)$, accompanied by the spin rotation operation $U_{C_{4z}}=e^{-i\frac{\pi}{4}\sigma_z}$ acting as $(\sigma_x,\sigma_y,\sigma_z)\to(\sigma_y,-\sigma_x,\sigma_z)$. Under this combined operation, the $d_{x^2-y^2}$-wave altermagnetic exchange term changes sign,
\begin{equation}
C_{4z}\,\Delta_{\mathrm{ALT}}(k_x,k_y)\sigma_z\,C_{4z}^{-1}
=-\Delta_{\mathrm{ALT}}(k_x,k_y)\sigma_z,
\end{equation}
whereas the RSOC term remains invariant. As a result, the static Hamiltonian does not possess $C_{4z}$ symmetry.

Despite the breaking of time-reversal and rotational symmetries individually, the static Hamiltonian preserves a combined antiunitary symmetry $\Theta C_4$, satisfying
\begin{equation}
(\Theta C_4)\,H^{\mathrm{static}}(\mathbf{k})\,(\Theta C_4)^{-1}
=H^{\mathrm{static}}(\mathbf{k}).
\end{equation}
This combined symmetry enforces vanishing net magnetization, even in the presence of momentum-dependent spin splitting.

\subsection{Effective Floquet Hamiltonian}

We now turn to the effective Floquet Hamiltonian generated by off-resonant circularly polarized light. The effective mass terms proportional to $J$ and $J'$ in $d_z^{\mathrm{eff}}(\mathbf{k})$ multiply $\sigma_z$ and are even functions of momentum. Consequently, time-reversal symmetry remains broken,
\begin{equation}
\Theta\,d_z^{\mathrm{eff}}(\mathbf{k})\sigma_z\,\Theta^{-1}
=-d_z^{\mathrm{eff}}(-\mathbf{k})\sigma_z.
\end{equation}
In contrast, both the static and light-induced spin--orbit coupling terms preserve time-reversal symmetry.

Regarding inversion, the scalar term $d_0^{\mathrm{eff}}$ and the mass term $d_z^{\mathrm{eff}}$ are even in momentum, whereas the in-plane components $d_x^{\mathrm{eff}}$ and $d_y^{\mathrm{eff}}$ are odd. Thus, inversion symmetry remains broken in the driven system, in direct analogy with the static RSOC case.

The fourfold rotational symmetry is also broken in the effective Floquet Hamiltonian. While the intrinsic altermagnetic exchange term changes sign under $C_{4z}$, the light-induced exchange contribution proportional to $J'$ is invariant. This mismatch explicitly violates $C_{4z}$ symmetry in the irradiated system.

Finally, since the light-induced spin-orbit coupling terms do not respect $C_{4z}$ symmetry, the combined $\Theta C_4$ symmetry that protected the static Hamiltonian is also broken in the effective Floquet Hamiltonian.
\section{Tight-Binding Floquet Hamiltonian}
\label{TB}
Let us reconstruct the effective Floquet Hamiltonian presented in Eq.~\ref{heff} in a more extended tight-binding form as,
\begin{equation}
\begin{aligned} 
H^{eff}(\textbf{k})=&\hat{T}_0+\hat{T}_x e^{ik_xa}+\hat{T}_y e^{ik_ya}+\\
&\hat{T}_{+x+y} e^{i(k_x+k_y) a}+\hat{T}_{-x+y} e^{i(-k_x+k_y)a}+ {\it c.c.}
\end{aligned}
\end{equation} 
where hopping matrices are defined as
\begin{equation}
\begin{aligned} 
\hat{T}_0=T_0,&\,\, \hat{T}_x=B T_x, \,\, \hat{T}_y=BT_y\\
\hat{T}_{\pm x+y}=&[i \lambda C(\sigma_x \pm \sigma_y)+ J^{\prime} \sigma_z]/4\\.
\end{aligned}
\end{equation}
Hopping matrices such as $T_0, T_x, T_y$ have already presented in Eq.~\ref{static_TB}. These tight-binding parameters are the basement for calculating the naoribbon band structures.

\section{Fourier Components of the Floquet Hamiltonian}
\label{FT}
To derive the effective Floquet Hamiltonian, the simplest way is to rewrite the $n^{th}$ Fourier component of $H^{(n)}$ represented in Eq.~\ref{hf}, for $n=0$, and also even and odd $n$'s. 
\begin{equation}
\begin{aligned}
H^{(0)}&=[4 t \sigma_0+M_z \sigma_z] [1-\mathcal{J}_0(a\mathcal{A}_0)]+H^{static}({\bf k})\mathcal{J}_0(a \mathcal{A}_0)\\
H^{(2m)}&=\mathcal{J}_{2m}(a\mathcal{A}_0)\Big\{ -2t[\cos(k_x a)+(-1)^m\cos(k_y a)]\sigma_0\\
&+2J[(-1)^m\cos(k_y a)-\cos(k_x a)]\sigma_z\\
&+\lambda [\sin(k_x a) \sigma_y -(-1)^m \sin(k_y a) \sigma_x] \Big\} \\
H^{(2m+1)}&=\mathcal{J}_{2m+1}(a\mathcal{A}_0)\Big\{ i [\lambda \cos(k_x a) \sigma_y\\
&+2(t \sigma_0+J\sigma_z) \sin(k_x a)]+(-1)^m[-\lambda \cos(k_y a) \sigma_x\\
&+2(t \sigma_0-J\sigma_z)\sin(k_y a)] \Big\} \\
\label{H_n}
\end{aligned}
\end{equation}
Regarding to the above formula and considering that $\mathcal{J}_{(-n)}(x)=(-1)^n\mathcal{J}_n(x)$, it is simply seen that $H^{(-2m)}=H^{(2m)}$, so the commutation relation for the even Fourier components would be zero,  $[H^{(-2m)},H^{(2m)}]=0$. 
\section{z-component Jacobian}
\label{Jacobian_app}
To determine the chirality of each Dirac points appeared in the BZ, it is neccasary to calculate the z-component Jacobian which is defined as, $J_3=(\partial_{k_x}{\vec d}\times\partial_{k_y}{\vec d})_z$. The sign of $J_3$ determines the chirality of that gapless point. For the effective Hamiltonian given in Eq.~\ref{heff}, this Jacobian reads as
\begin{equation}
\begin{aligned}
&J_3(\mathbf{k}) 
= (\lambda a)^2 \Big[
    C^2 \big( \sin(k_x a)\sin(k_y a) \big)^2 \\
   & + \cos(k_x a)\cos(k_y a)
      \big[ B + C\cos(k_x a) \big]
      \big[ B - C\cos(k_x a) \big]
\Big].\\
\end{aligned}\label{Jacobian_formula}
\end{equation}
As shown in the gap-closing conditions (Eq.~\ref{gap_closing_conditions}), one of the conditions is configured as ($\sin(k_y a)=0$ and $[B-C\cos(k_y a)]=0$) or ($\sin(k_x a)=0$ and $[B+C\cos(k_x a)]=0$) giving rise to $|B|=\pm|C|$. By looking at the Jacobian formula, Eq.~\ref{Jacobian_formula}, it is simply seen that $J_3$ is zero for these gap-closing conditions.

\section{Gap-Closing Conditions}
\label{gap_appendix}

In this appendix, we present the detailed gap-closing conditions employed in constructing the phase diagrams discussed in Sec.~\ref{S6}. Depending on the location of the band-touching points in momentum space, the gap-closing conditions can be classified into three distinct categories.

\begin{itemize}
    \item \textbf{(i) High-symmetry points.}  
    At the $\Gamma$ and $M$ points, the bulk gap closes when
    \[
    M_z + \frac{\lambda^2 C(x)}{2J} = 0 ,
    \]
    where $x = a\mathcal{A}_0$.  
    At the $X$ and $Y$ points, the gap-closing condition becomes
    \[
    M_z \pm 4J B(x) - \frac{\lambda^2 C(x)}{2J} = 0 ,
    \]
    with the plus (minus) sign corresponding to the $X$ ($Y$) point.  
    These conditions cannot be satisfied simultaneously; therefore, the $X$ and $Y$ points do not become gapless at the same parameter values.

    \item \textbf{(ii) Light-induced $G$ points.}  
    In addition to the high-symmetry points, the system hosts four symmetry-related gapless points $G_i$ ($i=1,\dots,4$), which satisfy
    \[
    \cos(k_y a) = -\cos(k_x a) = \frac{B}{C} .
    \]
    The bulk gap closes at these points when the effective mass term vanishes,
    \[
    M_z + \frac{4J B(x)^2}{C(x)}
    \left(1 - \frac{\lambda^2}{8J^2}\right) = 0 .
    \]
    This condition is realized only within the parameter window $|B| \le |C|$.

    \item \textbf{(iii) Marginal points.}  
    At the special parameter values $|B| = |C|$, additional gapless solutions occur at momenta $k_y = n\pi/a$ or $k_x = m\pi/a$. However, the associated Jacobian determinant vanishes at these points, implying zero chirality ($J_3=0$). Consequently, these marginal gapless points do not contribute to the Chern number and do not correspond to topological phase transitions.
\end{itemize}

\section{Absence of Anomalous Edge States}
\label{Pi_gap}
\begin{figure}
\includegraphics[width=\linewidth]{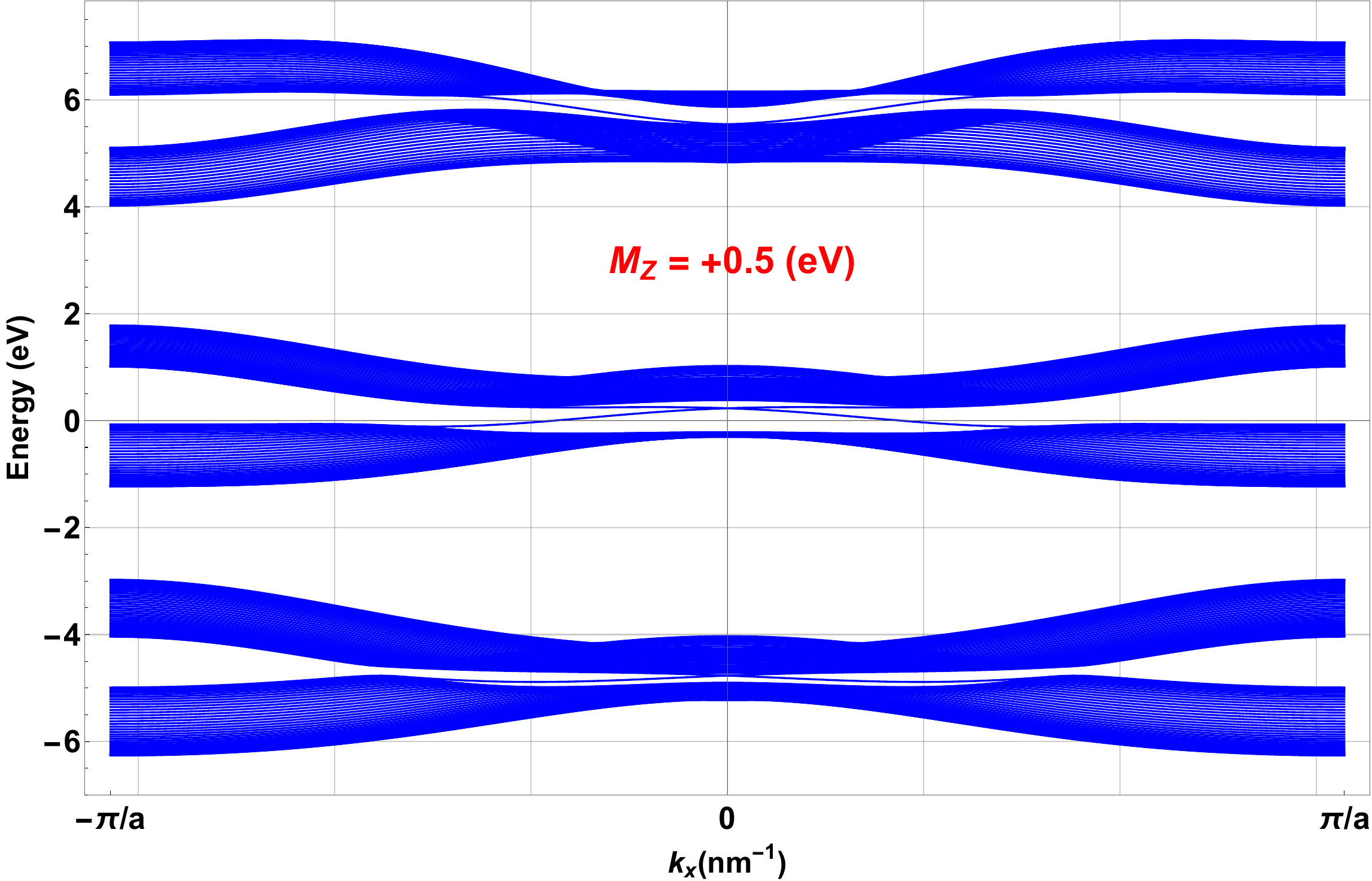} 
\caption{Floquet band spectrum of a nanoribbon of a $d_{x^2-y^2}$-wave altermagnet driven by right-handed circularly polarized light, showing the sidebands $n=0,\pm1$ for an out-of-plane magnetization $M_z=0.5~\mathrm{eV}$. The nanoribbon is oriented along the $x$ axis, and the energy spectrum is plotted as a function of $k_x$. Parameters are $J=1.0~\mathrm{eV}$, $\lambda=0.6~\mathrm{eV}$, $\Omega=5.0~\mathrm{eV}$, and the light amplitude $\mathcal{A}_0=0.38~\mathrm{nm^{-1}}$.
}
\label{Anomalous_edge}
\end{figure}
In the present work, we operate in the off-resonant, high-frequency regime,
where the driving frequency $\Omega$ is much larger than the electronic
bandwidth. In this limit, transitions between distant Floquet replicas are
strongly suppressed, and the quasienergy spectrum is dominated by the central
Floquet bands.

To explicitly verify the absence of anomalous Floquet edge modes, we computed
the full Floquet nanoribbon spectrum including the $n=0,\pm1$ sidebands.
The resulting quasienergy spectrum is shown in Fig.~\ref{Anomalous_edge}.
We observe that: (i) no chiral edge states appear inside the $\pi$ quasienergy gap, and
(ii) all edge modes are confined to the zero quasienergy gap.

This confirms that the driven altermagnet does not realize an anomalous Floquet
topological phase. Consequently, the topology of the system is fully
characterized by the Chern numbers of the effective Floquet Hamiltonian
$H^{\mathrm{eff}}$ derived in the high-frequency expansion.

\section*{Data Availability}
The data that support the findings of this study are available from the authors upon reasonable request.

\end{document}